\documentclass[prb,twocolumn,superscriptaddress,secnumarabic,balancelastpage,amsmath,amssymb,nofootinbib]{revtex4-1}
\usepackage{amssymb}
\usepackage{amsmath}
\usepackage{chngpage}
\usepackage{lgrind}        
\usepackage{color}         
\usepackage{graphics}      
\usepackage[pdftex]{graphicx}      
\usepackage{longtable}     
\usepackage{epsf}          
\usepackage{bm}            
\usepackage{asymptote}     
\usepackage{thumbpdf}
\usepackage{verbatim}
\usepackage{url}
\usepackage{MnSymbol}
\usepackage{physics}
\usepackage[colorlinks=true]{hyperref}  
\usepackage{enumitem}	
\usepackage{float}
\usepackage{times}
\usepackage{color}
\usepackage{verbatim}

\begin{document}


\title{Periodic table for topological bands with non-Hermitian Bernard-LeClair symmetries}


\author{Hengyun Zhou}
\email{hzhou@g.harvard.edu}
\author{Jong Yeon Lee}
\email{jlee12@g.harvard.edu}
\affiliation{Department of Physics, Harvard University, Cambridge, Massachusetts 02138, USA}


\begin{abstract}
Classifications of symmetry-protected topological (SPT) phases provide a framework to systematically understand the physical properties and potential applications of topological systems. While such classifications have been widely explored in the context of Hermitian systems, a complete understanding of the roles of more general non-Hermitian symmetries and their associated classification is still lacking. Here, we derive a periodic table for non-interacting SPTs with general non-Hermitian symmetries. Our analysis reveals novel non-Hermitian topological classes, while also naturally incorporating the entire classification of Hermitian systems as a special case of our scheme. Building on top of these results, we derive two independent generalizations of Kramers’ theorem to the non-Hermitian setting, which constrain the spectra of the system and lead to new topological invariants. To elucidate the physics behind the periodic table, we provide explicit examples of novel non-Hermitian topological invariants, focusing on the symmetry classes in zero, one and two dimensions with new topological classifications (e.g. $\mathbb{Z}$ in 0D, $\mathbb{Z}_2$ in 1D, 2D). These results thus provide a framework for the design and engineering of non-Hermitian symmetry-protected topological systems.
\end{abstract}
\maketitle

\section{Introduction}
Symmetry and topology lie at the heart of modern physics, and systematically understanding their roles in various physical systems has led to numerous interesting phenomena and potential applications \cite{hasan2010colloquium,moore2010birth,qi2011topological,chiu2016classification,armitage2018weyl,beenakker2016road,lu2016topological,huber2016topological,goldman2016topological,ozawa2018topological,wen2017colloquium}. Systematic classifications of symmetry-protected topological (SPT) phases, as exemplified by the ten-fold-way \cite{altland1997nonstandard,Kitaev2009,ryu2010topological} for free fermions, have been particularly important in providing a framework to analyze the topological behavior of systems with different symmetries and expediting the identification of new phases. Indeed, extensions of such approaches have also been proven to be extremely useful beyond the ten-fold-way, in the classification of topological crystalline insulators and gapless topological materials \cite{ando2015topological,fu2011topological,slager2013space,chiu2013classification,shiozaki2014topology,Kruthoff2017}. However, the majority of these works focus on the case of a closed Hermitian system.

In contrast, many physical systems, particularly in the context of atomic, molecular and optical physics
\cite{aspelmeyer2014cavity,cao2015dielectric,gao2015observation,doppler2016dynamically,peng2016anti,xu2016topological,zhen2015spawning}, may display richer non-Hermitian properties \cite{moiseyev2011non,feng2017non} associated with gain or loss in the system, leading to counter-intuitive phenomena such as unconventional transmission and reflection \cite{lin2011unidirectional,feng2013experimental}, parity-time symmetry \cite{bender1997real,guo2009observation,ruter2010observation,chong2011pt,konotop2016nonlinear,ambichl2013breaking,ge2014parity}, as well as laser and sensor applications \cite{hodaei2017enhanced,chen2017exceptional,lau2018non,zhang2018quantum,hodaei2014parity,feng2014single,el2014exceptional,liertzer2012pump,bandres2018topological,harari2018topological}. Moreover, it has recently been shown that even in solid-state systems conventionally described by a Hermitian Hamiltonian, an effective non-Hermitian description based on quasiparticle lifetimes can also yield new physical insights \cite{kozii2017non,shen2018quantum,papaj2018nodal,molina2018surface}. The topological properties of such non-Hermitian systems are also of great importance, both due to the fundamental interest of expanding the classes of available topological states \cite{alvarez2018topological,hatano1996localization,rudner2009topological,esaki2011edge,yuce2015topological,lee2016anomalous,leykam2017edge,leykam2017flat,xu2017weyl,shen2018topological,gong2018topological,qi2018defect,lieu2018topological,lieu2018topological1,alvarez2018topological,kawabata2018non,carlstrom2018exceptional,takata2018photonic,moors2018disorder,zyuzin2018flat,kunst2018non,zhou2018exceptional,budich2018symmetry,okugawa2018topological,dembowski2001experimental,poli2015selective,zeuner2015observation,weimann2016topologically,zhou2018observation,cerjan2018experimental,lapp2018engineering} 
and clarifying the roles of bulk-boundary correspondence in such systems \cite{xiong2018why,kunst2018biorthogonal,yao2018edge,yao2018non,lee2018anatomy}, but also due to their potential applications in e.g. topological lasers \cite{bandres2018topological,harari2018topological,st-jean2017lasing}.

However, systematic classifications of such non-Hermitian SPTs are still in progress. In a significant step towards this direction, Gong et al. \cite{gong2018topological} proposed an approach to extend the Hermitian classification techniques based on K-theory to non-Hermitian systems. However, only a limited set of symmetries that are directly realized in usual Hermitian systems were considered, and the classification was considered as an independent extension of Hermitian classes, where the usual Altland-Zirnbauer (AZ) classes were not directly included in the formalism. This calls for a systematic effort to analyze all possible symmetries in the non-Hermitian setting, and determine the allowed topological invariants.

Here, we systematically classify non-Hermitian topological bands in arbitrary spatial dimension, taking into account new types of symmetries that are unique to non-Hermitian systems. To this end, we make use of the Bernard-LeClair symmetry classes \cite{Bernard2001,bernard2001classification,magnea2008random,lieu2018topological}, based on four types of fundamental symmetries, which naturally generalize AZ classes into the case of non-Hermitian random matrix ensembles, resulting in a total of 38 symmetry classes. We find that in addition to expanding the classes of available symmetries, this approach also leads to two independent generalizations of Kramers' relations, which constrain the spectra and lead to degeneracies for certain symmetry classes, playing an important role in the identification of topological invariants. We then employ the technique of doubling the Hamiltonian (also known as Hermitian reduction) \cite{feinberg1997non,Bernard2001,gong2018topological,roy2017periodic} to reduce the non-Hermitian classification problem into a Hermitian classification problem, and apply K-theory techniques to obtain the classifying space and resultant topological invariants \cite{karoubi2006k}. Our classification scheme naturally includes previous results on non-Hermitian systems, and also contains Hermitian classifications as a special case, where Hermiticity is viewed as a special case of the more general pseudo-Hermiticity symmetry. To illustrate the periodic table, we analyze several nontrivial examples of the classification beyond what has been discussed in existing literature, providing detailed analysis of topological invariants in zero, one and two dimensions that did not appear in the limited symmetry classes of previous discussions of non-Hermitian topological phases. Our results thus provide a framework to analyze topological phases protected by general non-Hermitian symmetries, and could serve as an important guide for the experimental design of topologically-nontrivial non-Hermitian systems based on the novel invariants we have proposed.

This paper is organized as follows: In Sec.~\ref{sec:classifyherm}, we briefly review the Hermitian classification approaches based on K-theory and Clifford algebra extension problems. In Sec.~\ref{sec:symnonh}, we describe the Bernard-LeClair symmetry classes in detail, which form a natural generalization of the Altland-Zirnbauer classes to the non-Hermitian setting. Analyzing these symmetry classes in Sec.~\ref{sec:kramers}, we find two distinct non-Hermitian generalizations of Kramers degeneracy, one of which makes use of the biorthogonal properties of the system, and the other leading to spectra that forms complex-conjugate pairs. We then proceed to present our classification scheme and results in Sec.~\ref{sec:classifynonh}, which are based on a natural generalization of Hermitian gapped topological phases to non-Hermitian systems, culminating in the periodic table TABLE.~\ref{ref:tab0}. We find that in addition to reproducing known results, this new classification with more general non-Hermitian symmetries also provides new classes of topological invariants in non-Hermitian systems, and we elucidate their physical nature through explicit examples and calculations of topological invariants in Sec.~\ref{sec:newclass}, making use of symmetry transformations to Hermitian Hamiltonians, as well as correspondences between non-Hermitian Hamiltonians and block off-diagonal projectors of Hermitian Hamiltonians with chiral symmetry. We conclude by some remarks on different extensions to the classification scheme in Sec.~\ref{sec:discussion}. Details of some calculations as well as additional examples are presented in the appendices.

\section{Review of Hermitian Classifications}
\label{sec:classifyherm}
Before presenting our approach to the classification of SPTs protected by non-Hermitian symmetries, we first review methods to classify SPTs in Hermitian systems, based on K-theory and Clifford algebra extension problems \cite{altland1997nonstandard,Kitaev2009,ryu2010topological,stone2011symmetries,morimoto2013topological,chiu2014classification,chiu2016classification}. For simplicity, we shall be focusing on the case of internal symmetries only, although spatial symmetries can also be readily incorporated.

\subsection{Hermitian Symmetries}
The symmetry classes considered in the Hermitian setting (so-called AZ classes) are combinations of time-reversal symmetry $\mathcal{T}$, particle-hole symmetry $\mathcal{C}$, and chiral symmetry $\mathcal{P}$, defined as
\begin{align}
[H,\mathcal{T}]=\{H,\mathcal{C}\}=\{H,\mathcal{P}\}=0,
\label{eq:hermcomm}
\end{align}
where $H$ is the Hamiltonian, square brackets denote commutation and curly brackets denote anti-commutation. Each of these symmetries are involutions, meaning that acting with them twice gives rise to the same Hamiltonian. $\mathcal{T}$ and $\mathcal{C}$ are anti-unitary symmetries, while $\mathcal{P}$ is unitary. We assume that we are already operating in the symmetry sectors of any possible unitary, commuting symmetries such as spin-rotation symmetry, and can thus ignore their effects. Since the presence of two symmetries of the same kind will give rise to a unitary, commuting symmetry, we also assume that only one symmetry operator of each kind is present in this symmetry sector.

Since $\mathcal{P}$ is a unitary symmetry, we can multiply a phase such that it satisfies $\mathcal{P}^2=\mathbb{I}$. For anti-unitary symmetries, we can show that $\mathcal{T}^2,\mathcal{C}^2=\pm \mathbb{I}$ (see for example appendix~\ref{suppsec:symmetryclass}). Moreover, the combination of two of these symmetries will give rise to a symmetry of the other type.

This allows us to enumerate the AZ symmetry classes for Hermitian systems. When $H$ possesses anti-unitary symmetries, $\mathcal{T}$ and $\mathcal{C}$ can be either present (with square $\pm 1$) or not, giving rise to $3^2-1=8$ classes, where we do not count the case where both anti-unitary symmetries are not present. When $H$ does not possess any anti-unitary symmetries, there are 2 classes, depending on whether the chiral symmetry is present. Thus, there are a total of 10 possible symmetry classes.

\subsection{Clifford Algebra}
For classifications of Hermitian topological phases, Clifford algebras play an important role. These are algebras in which the generators anti-commute with each other. When there are anti-unitary symmetries present, the complex conjugation involved in the anti-unitary symmetry requires us to consider real Clifford algebras; otherwise, we consider complex Clifford algebras.

The complex Clifford algebra $Cl_n$, with $n$ generators $e_1,e_2,...,e_n$, is formed by taking all linear combinations of products of generators $e_1^{p_1}e_2^{p_2}...e_n^{p_n}$ ($p_i=0,1$) over the complex number field. In addition, the Clifford algebra generators satisfy the anti-commutation relation
\begin{align}
\{e_i,e_j\}=2\delta_{ij},
\end{align}
where we have used the fact that complex numbers can be multiplied to make each generator square to 1.

The real Clifford algebra $Cl_{p,q}$ has $n=p+q$ generators, and is formed by taking all linear combinations of products of generators $e_1^{p_1}e_2^{p_2}...e_n^{p_n}$ ($p_i=0,1$) over the \textit{real} number field. Since the underlying number field is real, we can no longer multiply arbitrary complex numbers to make the generators square to 1. Therefore, the generators are chosen to satisfy the following relations:
\begin{align}
\{e_i,e_j\}&=0,\quad j\neq i\\
e_i^2&=\begin{cases}
-1,& 1\leq i\leq p,\\
+1,& p+1\leq i\leq p+q.
\end{cases}
\end{align}

\subsection{Topological Classification for Hermitian Systems}
Having introduced the symmetries and mathematical language that will be employed in the classification, we can now proceed to discuss the topological classification of Hermitian systems, following Kitaev's approach \cite{Kitaev2009}. The results are summarized in TABLE.~\ref{tab:herm}.

First, we consider a zero-dimensional system. For a generic, gapped Hamiltonian, one first ``flattens" the spectra, keeping the gap open and thus preserving the topological properties, such that all eigenvalues above the gap are continuously deformed to lie at $+1$, and all eigenvalues below the gap lie at $-1$. The symmetry generators are then written in the form of matrix representations of Clifford algebra generators. For the classification to be generic and robust against the insertion of additional bands, as per Kitaev's original approach, the matrix representations should be chosen to be of sufficiently large dimension.

With the symmetry generators written as elements of a Clifford algebra $\{e_i\}$, the classification then corresponds to determining all possible inequivalent ways to insert the generator $e_0$, representing the mass term of the Hamiltonian, into the existing Clifford algebra. As an example with complex classes, when there are $n$ existing generators, this would correspond to the Clifford algebra extension problem $Cl_n\rightarrow Cl_{n+1}$. The set of such representations forms the so-called ``classifying space'', denoted $\mathcal{C}_q$ or $\mathcal{R}_q$ for the complex or real Clifford algebras.

According to K-theory, the classifying spaces for the complex and real Clifford algebra extension problems are
\begin{align}
Cl_n\rightarrow Cl_{n+1}\,\,\,\,\quad\Leftrightarrow&\quad \mathcal{C}_n,\\
Cl_{p,q}\rightarrow Cl_{p,q+1}\quad\Leftrightarrow&\quad\mathcal{R}_{q-p},\\
Cl_{p,q}\rightarrow Cl_{p+1,q}\quad\Leftrightarrow&\quad\mathcal{R}_{p+2-q}.
\end{align}

The distinct components are characterized by the zeroth homotopy group $\pi_0(C_q)$ or $\pi_0(R_q)$, which are well-known from the explicit forms of the classifying spaces based on the theory of symmetric spaces. Thus, one can determine the topological classification for any zero-dimensional system with the above symmetries.

A physical interpretation of this mathematical approach based on Clifford algebra extension problems is that we are seeking all inequivalent ways to insert a mass term and gap out a system, subject to some symmetry constraints. This interpretation also makes clear the classification approach in a general dimension $d$: we can generically consider a massive Dirac Hamiltonian
\begin{align}\label{eq:dirac_eqn}
H(\vec{k})=\sum_i k_i\gamma_i+m,
\end{align}
where $k_i$ is the momentum in the $i$th direction and $m$ is a mass term that gaps out the system. Similar to the zero-dimensional case, the mass term $m$ should satisfy the commutation relations Eq.~\ref{eq:hermcomm}. However, since anti-unitary operations flip the direction of momenta, the commutation relations with the $\gamma_i$ matrices become
\begin{align}
\{\mathcal{T},\gamma_i\}=[\mathcal{C},\gamma_i]=\{\mathcal{P},\gamma_i\}=0.
\end{align}
If we take the mass term $m$ to be flattened as in the 0D case, operators in Eq.~\ref{eq:dirac_eqn} should satisfy the commutation relations
\begin{align}
\{\gamma_i,\gamma_j\}=2\delta_{ij},\quad \{m,\gamma_i\}=0,\quad m^2=\mathbb{I}.
\end{align}
The topological classification then proceeds in a similar fashion as before. We classify the topologically inequivalent ways of adding the mass generator $e_0=m$ to the Clifford algebra formed by the symmetry generators \textit{and} the matrices $\gamma_i$. Thus, with each increase of dimension, the Clifford algebra extension problem and correspondingly the classifying space is shifted. This gives rise to the diagonal structure in the periodic table. Due to Bott periodicity in K-theory, the topological indices have period 2 for complex classes and period 8 for real classes. The classification results for Hermitian topological phases are summarized in TABLE.~\ref{tab:herm}.

\begin{table}
\begin{center}\begin{tabular}{ | c | c | c | c | c | c | c | c | c |}
\hline
Class & TRS & PHS & Chiral & Classifying space & 0D & 1D & 2D & 3D\\\hline
A & 0 & 0 & 0 & $\mathcal{C}_0$ & $\mathbb{Z}$ & 0 &  $\mathbb{Z}$ & 0\\
AIII & 0 & 0 & 1 & $\mathcal{C}_1$ & 0 & $\mathbb{Z}$ & 0 & $\mathbb{Z}$\\
AI & 1 & 0 & 0 & $\mathcal{R}_0$ & $\mathbb{Z}$ & 0 & 0 & 0\\
BDI & 1 & 1 & 1 & $\mathcal{R}_1$ & $\mathbb{Z}_2$ & $\mathbb{Z}$ & 0 & 0\\
D & 0 & 1 & 0 & $\mathcal{R}_2$ & $\mathbb{Z}_2$ & $\mathbb{Z}_2$ & $\mathbb{Z}$ & 0\\
DIII & -1 & 1 & 1 & $\mathcal{R}_3$ & 0 & $\mathbb{Z}_2$ & $\mathbb{Z}_2$ & $\mathbb{Z}$\\
AII & -1 & 0 & 0 & $\mathcal{R}_4$ & $\mathbb{Z}$ & 0 & $\mathbb{Z}_2$ & $\mathbb{Z}_2$\\
CII & -1 & -1 & 1 & $\mathcal{R}_5$ & 0 & $\mathbb{Z}$ & 0 & $\mathbb{Z}_2$\\
C & 0 & -1 & 0 &  $\mathcal{R}_6$ & 0 & 0 & $\mathbb{Z}$ & 0\\
CI & 1 & -1 & 1 & $\mathcal{R}_7$ & 0 & 0 & 0 & $\mathbb{Z}$\\\hline
\end{tabular}
\end{center}
\caption{Periodic table for Hermitian topological phases, adapted from Ref.~\onlinecite{Kitaev2009,ryu2010topological,morimoto2013topological}. The columns specify the symmetry class, the square of each symmetry (time-reversal symmetry, particle-hole symmetry, and chiral symmetry), where 0 denotes that this symmetry is  not present, the classifying space for the zero-dimensional Clifford algebra extension problem, and the topological classification in dimensions 0 to 3. \label{tab:herm}}
\end{table}

\section{Non-Hermitian Symmetry Classes}
\label{sec:symnonh}
We now generalize the symmetry classes to the non-Hermitian case, making use of the ideas of Bernard-LeClair symmetry classes
\cite{Bernard2001,bernard2001classification,magnea2008random,sato2012time,lieu2018topological}. The key difference is that in the case of non-Hermitian systems, the scope of symmetries is significantly expanded; in particular, Hermiticity can now be viewed as a special type of non-Hermitian symmetry (type $Q$). Time-reversal symmetry and particle-hole symmetry become equivalent under an imaginary rotation \cite{kawabata2018topological}, but can have two independent types (type $C$ and type $K$) of generalizations to the non-Hermitian setting. Combined with chiral symmetries (denoted type $P$), this gives rise to four different types of symmetries, with certain equivalence relations between different combinations of them.

\subsection{Basic Building Blocks: Bernard-LeClair Symmetry Classes}
\label{sec:symmetrybasic}
We first explain the form of the basic symmetries. For unitary, commuting symmetries, the Hamiltonian can be block-diagonalized into different symmetry sectors and considered separately. Therefore, we focus on the remaining possible symmetries, restricted on physical grounds to be involutions, that act on each symmetry sector. From a physics viewpoint, we would like them to be natural---but complete---generalizations of the Hermitian time-reversal, particle-hole, and chiral symmetries. Moreover, we would like to directly incorporate Hermitian classifications into this formalism. This motivates the following forms of symmetries that we consider.

Generically, we may write
\begin{align}
h = \epsilon_{\mathcal{O}} u\mathcal{O}(h)u^{-1},
\end{align}
where the operation $\mathcal{O}$ can be identity, Hermitian conjugation, transposition or complex conjugation, labeled as type  $P, Q, C, K$ symmetries, and $\epsilon_{\mathcal{O}}$ is a sign factor $\pm 1$, as required for an involution. Since we are already in a fixed symmetry sector, we assume no unitary, commuting symmetries, and therefore must have $\epsilon_p=-1$, in analogy to the Hermitian chiral symmetry. Type $P$ and $Q$ symmetries are straightforward generalizations of chiral symmetry and Hermiticity, while type $C$ and $K$ are two different generalizations of time-reversal/particle-hole symmetry that coincide in the Hermitian case but can generically be different in the non-Hermitian setting.

We note that the presence of some symmetries may imply others. For example, since each of the symmetries are restricted to be involutions, a combination of two symmetries (specified by $u_1$, $u_2$) of the same type will result in a unitary commuting symmetry $u_1u_2$, and thus within each symmetry sector, we only need to consider at most a single instance of each symmetry. Moreover, the combination of a type $Q$ and type $C$ symmetry automatically implies a type $K$ symmetry $k=qc$; therefore, to enumerate all classes that involve any two or three of type $Q$, $C$, $K$ symmetry, we only need to consider the inclusion of a type $Q$ and a type $C$ symmetry.

For the two symmetry types $Q$ and $K$ that involve complex conjugation, we may redefine $h\rightarrow ih$ to flip the sign of $\epsilon_{\mathcal{O}}$, and thus without loss of generality we may choose $\epsilon_q=\epsilon_k=1$. Meanwhile, $\epsilon_c=\pm 1$ can take on either sign. Note that care must be taken when both type $Q$ and type $K$ symmetries are present, since the above redefinitions to make $\epsilon_q=\epsilon_k=1$ may be inconsistent. However, as discussed above, any two of type $Q$, $C$, $K$ symmetries will imply the remaining one; therefore, we can always consider only the corresponding type $Q$ and type $C$ symmetries in this scenario, and use the consistent sign for the type $K$ symmetry that is automatically implied. The choice of sign will impact the spectrum, but will not modify the topological properties that we are interested in.

The condition that the symmetries are involutions also imposes restrictions on the unitary symmetry implementation $u$. As is shown in Appendix~\ref{suppsec:symmetryclass}, the unitary implementations $p$, $q$, $c$, $k$ of type $P$, $Q$, $C$, $K$ symmetries are required to satisfy $p^2=q^2=cc^*=kk^*=\lambda \mathbb{I}$. Moreover, by an appropriate phase choice of $p$ and $q$, and by further analysis of $c$ and $k$, we find that we can choose $p^2=q^2=\mathbb{I}$ and $cc^*=kk^*=\pm \mathbb{I}$.

Combining all of the preceding considerations, we arrive at the canonical forms for the 4 types of symmetries:
\begin{align}
h=-php^{-1}&, p^2=\mathbb{I}&(P \textrm{ sym.})\label{eq:syms1}\\
h=qh^\dagger q^{-1}&, q^2=\mathbb{I}&(Q \textrm{ sym.})\label{eq:syms2}\\
h=\epsilon_c ch^Tc^{-1}&, cc^*=\eta_c \mathbb{I}&(C \textrm{ sym.})\label{eq:syms3}\\
h=kh^*k^{-1}&, kk^*=\eta_k \mathbb{I}&(K \textrm{ sym.})
\label{eq:syms4}
\end{align}

For physical dimensions greater than 0, we also need to specify how the symmetries operate on the momentum. Since type $P$ and $Q$ symmetries are more analogous to Hermiticity and chiral symmetry, we expect them to preserve the momentum $k\rightarrow k$, while for type $C$ and $K$ symmetries, since they are more analogous to time-reversal symmetry and particle-hole symmetry, we expect them to behave similar to anti-linear and anti-unitary symmetries, and should thus bring $k\rightarrow -k$. This is further justified in Sec.~\ref{sec:doubling}, where we discuss the use of doubled Hamiltonians to perform topological classifications; type $P$, $Q$ symmetries are mapped to unitary symmetries at the doubled level, while type $C$, $K$ symmetries are mapped to anti-unitary symmetries, consistent with the above choice of action on the momentum. This is consistent with the AZ symmetry classes, but for other physical situations, e.g. $\mathcal{PT}$ symmetry, it may be desirable to adopt different momentum transformation rules.

\subsection{Commutation Relations}
To consider the combination of several symmetries, we require that the different transformations specified in Eqs.~\ref{eq:syms1}-\ref{eq:syms4} commute. Taking type $P$ and $Q$ symmetries as an example, this implies
\begin{align}
    -pqh^\dagger q^\dagger p^\dagger=-qph^\dagger p^\dagger q^\dagger
\end{align}
for generic $h$, which implies that $p^\dagger q^\dagger pq=\lambda \mathbb{I}$, or in other  words $\lambda p=q^\dagger pq$. Taking the square, and using the fact that we can choose $p^2=\mathbb{I}$, we find that $\lambda^2=1$. Thus, we find that
\begin{align}
p=\epsilon_{pq}q^\dagger pq,\quad \epsilon_{pq}=\pm 1.
\end{align}
Similarly, we find that
\begin{align}
c=\epsilon_{pc}pcp^T,\quad k=\epsilon_{pk}pkp^T,\quad c=\epsilon_{qc}qcq^T,
\label{eq:commute}
\end{align}
where $\epsilon_{\mu\nu}=\pm 1$ for $\mu,\nu=p,q,c,k$.

The types of symmetries considered, combined with the signs of the commutation relation $\epsilon_{\mu\nu}$,  the sign of the commutation between the type $C$ symmetry and the Hamiltonian $\epsilon_c$, and the sign of the symmetry operator involution identity $\eta_c$, $\eta_k$, completely specify a symmetry class. However, some of these symmetry classes can be shown to be equivalent to each other, as we now discuss.

\subsection{Equivalence Relations}
\label{sec:equivrel}
Similar to how a type $Q$ and a type $C$ symmetry may combine to imply a type $K$ symmetry, combinations of a type $P$ symmetry and another symmetry may also imply a new symmetry of the second kind that satisfies a different commutation relation from the original. This leads to equivalence relations between different possible commutation relations and operator identities. Here, we provide a detailed derivation of such relations.

First, consider a type $P$ and type $C$ symmetry. This implies that there exists an additional type $C$ symmetry implemented as $\tilde{c}=pc$. Suppose that the original symmetries have parameters for the square and commutation relations as specified above, then the corresponding properties for the new type $C$ symmetry implemented by $\tilde{c}$ can be found to be:
\begin{eqnarray}
h=-php^{-1}=-\epsilon_c \tilde{c}h^T\tilde{c}^{-1} &\Rightarrow& \epsilon_{\tilde{c}} = -\epsilon_c \quad\,\,\,\\
\tilde{c}\tilde{c}^*=pcp^*c^*=pcp^Tc^T=\epsilon_{cp}cc^*=\epsilon_{cp}\eta_c &\Rightarrow& \eta_{\tilde{c}}=\epsilon_{cp}\eta_c \quad\,\,\,\\
p\tilde{c}p^T=cp^T=\epsilon_{cp}pc=\epsilon_{cp}\tilde{c} &\Rightarrow& \epsilon_{\tilde{c}p}=\epsilon_{cp}\quad\,\,\,\\
\tilde{c}^\dagger q^\dagger \tilde{c}=c^\dagger p^\dagger q^\dagger pc=\epsilon_{pq}c^\dagger q^\dagger c=\epsilon_{pq}q^\dagger &\Rightarrow& \epsilon_{q\tilde{c}}=\epsilon_{qp}\epsilon_{qc}.\quad\,\,\,
\end{eqnarray}
Similarly, for a type $P$ and type $K$ symmetry, we shall find that an additional type $K$ symmetry $\tilde{k}=pk$ is implied, and that this symmetry satisfies the condition $\eta_{\tilde{k}}=\epsilon_{pk}\eta_k$, $\epsilon_{p\tilde{k}}=\epsilon_{pk}$. Note that in this calculation, we have already used the fact, as discussed previously, that for the type $K$ symmetry, we may redefine the Hamiltonian as $h\rightarrow ih$ to get rid of any signs in front of the symmetry definition.

Finally, we note that combining a type  $P$ and type $Q$ symmetry also leads to an additional type $Q$ symmetry implemented as $\tilde{q}=qp$. In order to satisfy the square condition for a type $Q$ symmetry $\tilde{q}^2=\mathbb{I}$, we should multiply an additional phase factor
\begin{align}
    \tilde{q}=\sqrt{\epsilon_{pq}}qp,
\end{align}
since $qpqp=\epsilon_{pq}$. The only commutation relation that can be modified by this redefinition is between the type $Q$ symmetry and a type $C$ symmetry. In particular, using the identities in Eq.~\ref{eq:commute}, we find that
\begin{align}
    c\tilde{q}^*&=(\sqrt{\epsilon_{pq}})^*cq^*p^*\nonumber\\&=(\sqrt{\epsilon_{pq}})^*\epsilon_{qc}\epsilon_{pc}qpc=\epsilon_{qc}\epsilon_{pc}\epsilon_{pq}\tilde{q}c,
    \label{eq:p+q}
\end{align}
so that $\epsilon_{\tilde{q}c}=\epsilon_{qc}\epsilon_{pc}\epsilon_{pq}$.

By enumerating all possible symmetries, signs of the symmetry relations, and their equivalence relations, we obtain a total of 38 symmetry classes. Note that this is slightly different from the number of classes noted in Ref.~\onlinecite{Bernard2001,magnea2008random}, since in those works inequivalent representations of the same symmetry were separated into different classes, and in Ref.~\onlinecite{Bernard2001} several symmetry classes were accidentally dropped. However, since we are here interested in the topological classification resulting from the symmetry classes, we are agnostic to the specific unitary implementation of the symmetry, and regard different representations of the symmetry as belonging to the same class. Note however that even though inequivalent representations of the same symmetry will be shown to have an identical topological classification, the precise nature of the invariant may still be slightly different, as will be discussed in Sec.~\ref{sec:newclass}.

\section{Non-Hermitian Kramers Degeneracy}
\label{sec:kramers}
Before providing the full periodic table for SPTs based on these non-Hermitian symmetry classes, we first comment on some physical consequences of the non-Hermitian symmetries, which will also play an important role in understanding the nature of topological invariants for certain models. In particular, we prove the non-Hermitian counterpart of Kramers relations for type $C$ and type $K$ symmetries, when the symmetries square to -1. In the Hermitian limit, both of these cases reduce to the Hermitian time-reversal symmetry. However, in the non-Hermitian case, the form of the Kramers relation will be considerably different between the two types of symmetries. Note that our proof does not rely on pseudo-Hermiticity, in contrast with previous studies of generalized Kramers degeneracies in non-Hermitian systems \cite{sato2012time}, where the system has both type $K$ and $Q$ symmetries; in those cases, under the presence of a type $Q$ symmetry, the problem can be directly transformed to the usual Kramers degeneracy via the method described in Sec.~\ref{sec:altq}.

\subsection{Generalized Kramers Relation for Type $K$ Symmetry}
We first prove a generalized Kramers relation for type $K$ symmetries with $\eta_k=kk^*=-1$, which has a closer resemblance to the proof of Kramers degeneracy in Hermitian systems. We will show that as a consequence of the type $K$ symmetry, each eigenstate has an associated pair with a complex-conjugated eigenvalue. Thus, in this case, the symmetry does not impose a full degeneracy of the eigenvalue, but only guarantees that the real part of the eigenvalues are degenerate.

Using the right eigenvalue equation, we find that
\begin{align}
    kh^*_{-\vec{k}}k^\dagger v_{\vec{k}}&=h_{\vec{k}}v_{\vec{k}}=\lambda_{\vec{k}} v_{\vec{k}},\nonumber\\
    \Rightarrow\quad h_{-\vec{k}}\qty(k^Tv^*_{\vec{k}})&=\lambda^*_{\vec{k}}(k^Tv^*_{\vec{k}})
\end{align}
where $v_{\vec{k}}$, $\lambda_{\vec{k}}$ are a pair of right eigenvector and eigenvalue. Thus, we find that $k^Tv^*_{\vec{k}}$ and $\lambda^*_{\vec{k}}$ are a pair of right-eigenvector and eigenvalue of $h_{-\vec{k}}$. This implies that the spectra forms complex conjugate pairs between $\vec{k}$ and $-\vec{k}$.

At the time-reversal invariant points $\vec{k}=-\vec{k}$, we need to show that the resulting eigenvectors from this symmetry operation $v_{\vec{k}}$ and $k^Tv^*_{\vec{k}}$ are independent of each other. Suppose they are linearly dependent, then
\begin{equation}
    v_{\vec{k}} = e^{i\phi} k^T v^*_{\vec{k}} = e^{i\phi}k^Te^{-i\phi}k^\dagger v_{\vec{k}} = k^T k^\dagger  v_{\vec{k}},
\end{equation}
which contradicts with the fact that $\eta_k=kk^*=k^Tk^\dagger=-1$. Therefore, the generalized Kramers relation holds, and each eigenstate has an associated pair with a complex-conjugated eigenvalue. In the case where $\eta_k=kk^*=+1$, this relation will only hold for complex eigenvalues, and will not hold anymore for real eigenvalues at the time-reversal invariant points, in analogy to the fact that Kramers degeneracy holds for fermionic systems, but not bosonic systems. This result is consistent with the $2\mathbb{Z}$ invariant found in Ref.~\onlinecite{gong2018topological} for non-Hermitian class AII/C (see also class 35 in TABLE.~\ref{ref:tab2}).

\subsection{Biorthogonal Kramers Degeneracy for Type $C$ Symmetry}
\label{sec:typeckramers}
We now prove a generalization of Kramers degeneracy for the type $C$ symmetry with $\eta_c=cc^*=-1$, in which the symmetry is sufficient to guarantee that for each eigenstate, there is an associated pair with the same complex eigenvalue. This is directly applied in Sec.~\ref{sec:1D}, where the type $C$ symmetry provides a robust two-fold degeneracy for a Hamiltonian in class 7. Since the type $C$ symmetry makes use of a transpose operation, it is necessary to consider both left and right eigenvectors.

To proceed, let us write left and right eigenvalue equations as $v_L h= \lambda_L v_L$ and $h v_R  = \lambda_R v_R$; it is well-known that the set of left and right eigenvalues must be the same, since both of them are roots of the determinant equation $\det(h - \lambda \mathbb{I}) = 0$. However, since $h \neq h^\dagger$, left and right eigenvectors are no longer related by a conjugate transpose, implying that left (right) eigenvectors are not orthonormal among themselves anymore. The correct generalization of orthonormality is that the left and right eigenvectors form a biorthogonal system, where $v_L^n v_R^m=\delta_{n,m}$, corresponding  to the unconjugated inner product between the row ($v_L^n$) and column ($v_R^m$) vectors being a kronecker delta.

In the presence of a type $C$ symmetry, one can relate left and right eigenvectors. Since this symmetry maps $\vec{k}$ to $-\vec{k}$, $h_{-\vec{k}} = c h_{\vec{k}}^T c^\dagger$, 
\begin{eqnarray}
\lambda_{L,\vec{k}} v_L = v_L h_{\vec{k}} &=& v_L c h_{-\vec{k}}^T c^\dagger \nonumber \\
\Rightarrow \;\;\quad \lambda_{L,\vec{k}} (v_{L,\vec{k}} c) &=& (v_{L,\vec{k}} c) h_{-\vec{k}}^T \nonumber \\
\Rightarrow \quad \lambda_{L,\vec{k}} (v_{L,\vec{k}} c)^T &=& h_{-\vec{k}} (v_{L,\vec{k}} c)^T
\end{eqnarray}
Therefore, if $v_{L,\vec{k}}$ is a left eigenvector of $h_{\vec{k}}$, then $(v_{L,\vec{k}}c)^T$ is a right eigenvector of $h_{-\vec{k}}$.

To show the generalized Kramers degeneracy, it is necessary to show that at the time-reversal invariant points $\vec{k}=-\vec{k}$, the resulting right eigenvector $(v_L^nc)^T$ is different from the biorthonormal partner of $v_L^n$. To proceed, we observe that $\eta_c=c c^* = -1$ implies that $c^T = - c$. Now, suppressing the momentum index for the time-reversal invariant points, let us write $v_L (v_L c)^T = \lambda$, then
\begin{eqnarray}
    \lambda &=& v_L (v_L c)^T = v_L c^T v_L^T  \nonumber\\
    &=& (v_L c^T v_L^T)^T \quad \text{ as it is a number} \nonumber \\
    &=& v_L c v_L^T = -v_L c^T v_L^T = -\lambda,
\end{eqnarray}
implying that $\lambda = 0$. Thus, $v_L$ and $(v_Lc)^T$ are orthogonal, and do not form a biorthonormal pair. This implies that at the time-reversal invariant point, there exists a pair of independent right (left) eigenvectors with the same eigenvalue, thus ensuring the generalized Kramers degeneracy. Combined with the relation between eigenvalues at $h_{\vec{k}}$ and $h_{-\vec{k}}$ described above, we also find that the winding direction of two bands forming a Kramers pair will be opposite from each other in 1D. This crucial consequence on the spectra will play an important role in our determination of the $\mathbb{Z}_2$-topological invariant in Sec.~\ref{sec:1D}.

\section{Topological Classification for  Non-Hermitian Systems}
\label{sec:classifynonh}

\begin{figure}
    \centering
     \includegraphics[width=0.49\textwidth]{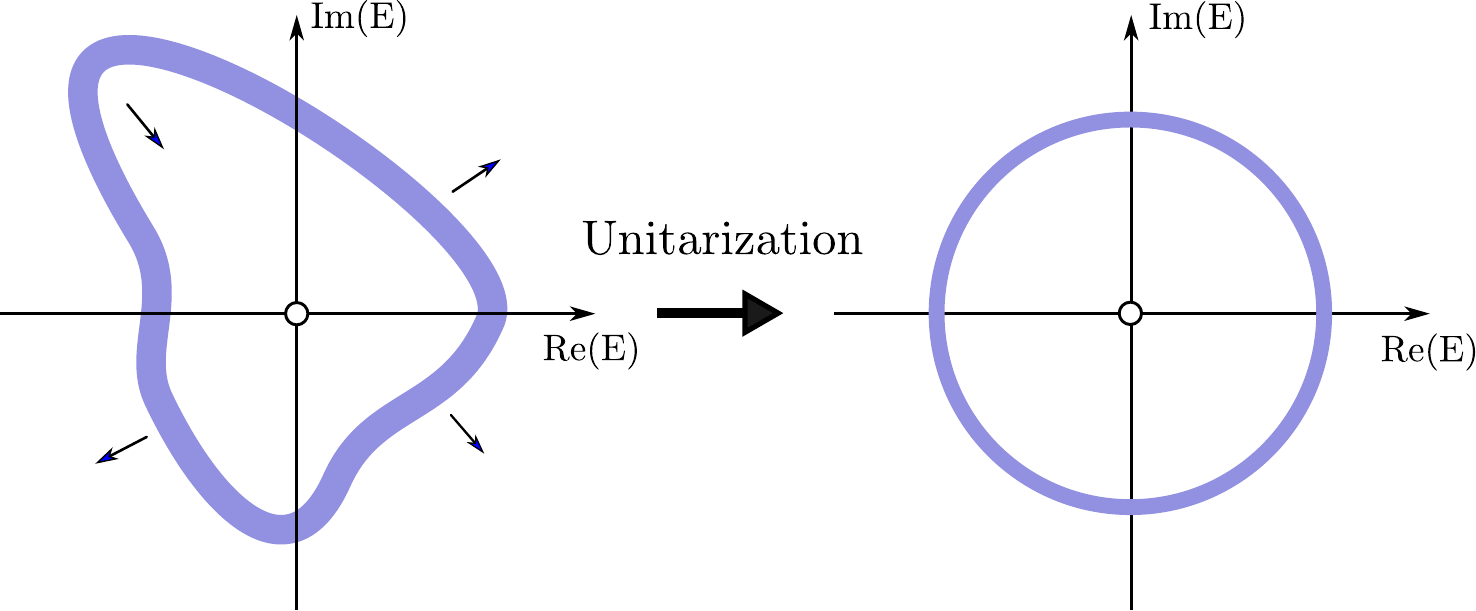}
    \caption{In the non-Hermitian setting, two configurations are equivalent if one can deform the complex spectra to another, subject to symmetry constraints, while not touching the base energy $E=0$ (origin in this figure). By this deformation, one can transform the Hamiltonian into a unitary matrix.
    }
    \label{fig:NH_gap}
\end{figure}
Equipped with an understanding of the non-Hermitian symmetry classes, we now proceed to classify the topological band structure for each symmetry class. Before providing the full classification, however, we first clarify the type of topology that we are considering here. 

In Hermitian systems, the topology of the system is provided by equivalence classes of Hamiltonians, subject to desired symmetry conditions, where Hamiltonians belonging to different classes cannot be continuously deformed into each other without closing the band gap \cite{Kitaev2009}. Another way to view this condition is that the energy never crosses a specified value, which is typically chosen to be the Fermi energy $E_F$ and shifted to be set to 0.

In the non-Hermitian setting, the spectra can now be complex, and correspondingly one natural generalization of the gapped condition is for the spectrum to not touch a complex base energy $E$ \cite{gong2018topological}, which we shift to $0$ without loss of generality (see Fig.~\ref{fig:NH_gap}). This can be viewed as a dynamical generalization of the gapped condition, where in addition to conditions on the real part of the energy, the dynamical influence of gain and loss are also important. As we shall see, such a condition also allows us to directly establish a mapping to Hermitian classification problems.

In the following, we discuss in detail the procedure for deriving the classifications for SPTs protected by non-Hermitian symmetries. We first perform a unitarization procedure, in analogy to band-flattening in the Hermitian case, in order to bring the non-Hermitian Hamiltonian into a standard form. The unitarized Hamiltonian is then doubled into a Hermitian Hamiltonian with an additional chiral symmetry, for which a one-to-one correspondence between topological classes can be specified. We then proceed to provide the classification for the corresponding Hermitian problem, and use this to obtain the classifications for SPTs protected by non-Hermitian symmetries.

\subsection{Unitarization of Non-Hermitian Hamiltonians}
In analogy to the flattening procedure in Hermitian classification approaches \cite{Kitaev2009}, where the spectrum is deformed to have eigenvalues of $\pm1$, here we perform a unitarization procedure \cite{gong2018topological}, such that the eigenvalues are brought onto a unit circle centered at the base reference point, which is here chosen to be 0. This provides a convenient canonical form for the following classifications.

More specifically, we show that any invertible Hamiltonian $h$ can be continuously deformed (in a symmetry-respecting way) into a unitary matrix $u$, where $u$ is given by the unitary matrix in the polar decomposition of $h=uP$, $P=\sqrt{h^\dagger h}$ is positive-definite. This decomposition will always be unique as long as the eigenvalue spectrum does not touch the base point (i.e. it is invertible), which is our prescribed condition for equivalence classes of Hamiltonians.

To prove the existence of a continuous deformation from $h$ to $u$, we utilize the following property of positive-definite matrices: if $A$ and $B$ are positive-definite matrices, and $A^2=B^2$, then $A=B$, since a positive-definite matrix has a unique positive-definite square root. Using this property, we show in appendix \ref{suppsec:unitarization} that $u$ respects the same non-Hermitian symmetries as $h$. This implies that the path $h(\lambda)=\lambda h+(1-\lambda)u=u[\lambda P+1-\lambda]$, for $\lambda$ going from 0 to 1, will be a symmetry-respecting continuous path connecting the non-Hermitian Hamiltonian $h$ and the unitary matrix $u$ defined above, along which the Hamiltonian remains invertible. Thus, symmetry-respecting unitarization is always possible.

\subsection{Doubling the Hamiltonian}
\label{sec:doubling}
With the non-Hermitian Hamiltonian $h$ now deformed into a unitary Hamiltonian $u$, we can make use of previous classification results of Floquet SPTs \cite{roy2017periodic} by rewriting the unitary matrix as a doubled Hamiltonian \cite{Bernard2001,gong2018topological,roy2017periodic}
\begin{align}
H_u=\begin{pmatrix}
0 & u \\  u^\dagger  &  0
\end{pmatrix},
\end{align}
which by construction is guaranteed to be Hermitian, squares to identity, and automatically possesses a chiral symmetry of the form $\Sigma=\sigma_z\otimes\mathbb{I}$. As has been shown in Ref.~\onlinecite{gong2018topological,roy2017periodic}, there is a one-to-one correspondence between the homotopy classes of $u$ with the corresponding ones of $H_u$, where $H_u$ is subject to the constraints of the chiral symmetry $\Sigma$, Hermiticity, and a gapped spectrum. This thus allows us to obtain the topological classification of a non-Hermitian system from its Hermitian counterpart, for which mature techniques have been developed.

To obtain the corresponding Hermitian classification problem, we also need to understand how the symmetries are mapped over to the case of the doubled Hamiltonian $H_u$. Defining
\begin{align}
\bar{P}=\begin{pmatrix}
p & 0\\ 0 & p
\end{pmatrix},\quad 
\bar{Q}=\begin{pmatrix}
0 & q\\ q & 0
\end{pmatrix},\quad
\bar{C}=\begin{pmatrix}
0 & c\\
c & 0\\
\end{pmatrix}\mathcal{K},\quad
\bar{K}=\begin{pmatrix}
k & 0\\
0 & k
\end{pmatrix}\mathcal{K},
\label{eq:doubsym}
\end{align}
where $\mathcal{K}$ denotes complex conjugation, we find that the original symmetry definitions imply for the doubled Hamiltonians that
\begin{align}
\qty{\bar{P},H_u}=\qty[\bar{Q},H_u]=\qty[\bar{K},H_u]=0,\, H_u=\epsilon_c\bar{C}H_u\bar{C}^{-1}.
\end{align}
Note that the last two ``doubled'' symmetries $\bar{C}$ and $\bar{K}$ are actually anti-unitary and we prescribed them to send momentum $\vec{k} \mapsto -\vec{k}$. In addition, we can directly verify that
\begin{align}
&\qty[\bar{P},\Sigma]=\qty[\bar{K},\Sigma]=\qty{\bar{Q},\Sigma}=\qty{\bar{C},\Sigma}=0,\\
&\bar{P}^2=\bar{Q}^2=\eta_c\bar{C}^2=\eta_k\bar{K}^2=\mathbb{I},
\end{align}
and the commutation relations between the different types of symmetries carry over to the doubled Hamiltonian.

Thus, working with the doubled Hamiltonian $H_u$, the inherent chiral symmetry $\Sigma$, and the doubled symmetries $\bar{P}$, $\bar{Q}$, $\bar{C}$, $\bar{K}$, and having specified their commutation relations and squares, we can employ standard K-theory techniques to determine the classifying space and resulting topological classification. In the following, we clarify a few technical points that arise when performing this procedure.

\begin{table*}[!htbp]
\resizebox{2.05\columnwidth}{!}
{\begin{tabular}{ | c | c | c | c | c | c | c | c | c | c | c | c | c | c |}\hline Sym. & Gen. Rel. & Cl. & $d = 0$ & 1 & 2 & 3 & Sym. & Gen. Rel. & Cl. & $d = 0$ & 1 & 2 & 3\\\hline1,  &  &  $\mathcal{C}_1$ & $0$ & $\mathbb{Z}$ & $0$ & $\mathbb{Z}$ & 20, QC & $\epsilon_c=-1$, $\eta_c=1$, $\epsilon_{qc}=-1$ &  $\mathcal{R}_4$ & $\mathbb{Z}$ & $0$ & $\mathbb{Z}_2$ & $\mathbb{Z}_2$\\\hline2, P &  &  $\mathcal{C}_1^{\times 2}$ & $0$ & $\mathbb{Z}^{\times 2}$ & $0$ & $\mathbb{Z}^{\times 2}$ & 21, QC & $\epsilon_c=-1$, $\eta_c=-1$, $\epsilon_{qc}=-1$ &  $\mathcal{R}_0$ & $\mathbb{Z}$ & $0$ & $0$ & $0$\\\hline3, Q &  &  $\mathcal{C}_0$ & $\mathbb{Z}$ & $0$ & $\mathbb{Z}$ & $0$ & 22, PQC & \begin{tabular}{c}$\epsilon_c=1$, $\eta_c=1$, $\epsilon_{pq}=1$, $\epsilon_{pc}=1$, $\epsilon_{qc}=1$\\\hline$\epsilon_c=-1$, $\eta_c=1$, $\epsilon_{pq}=1$, $\epsilon_{pc}=1$, $\epsilon_{qc}=1$\end{tabular} &  $\mathcal{R}_1$ & $\mathbb{Z}_2$ & $\mathbb{Z}$ & $0$ & $0$\\\hline4, PQ & $\epsilon_{pq}=1$ &  $\mathcal{C}_1$ & $0$ & $\mathbb{Z}$ & $0$ & $\mathbb{Z}$ & 23, PQC & \begin{tabular}{c}$\epsilon_c=1$, $\eta_c=-1$, $\epsilon_{pq}=1$, $\epsilon_{pc}=1$, $\epsilon_{qc}=1$\\\hline$\epsilon_c=-1$, $\eta_c=-1$, $\epsilon_{pq}=1$, $\epsilon_{pc}=1$, $\epsilon_{qc}=1$\end{tabular} &  $\mathcal{R}_5$ & $0$ & $\mathbb{Z}$ & $0$ & $\mathbb{Z}_2$\\\hline5, PQ & $\epsilon_{pq}=-1$ &  $\mathcal{C}_0^{\times 2}$ & $\mathbb{Z}^{\times 2}$ & $0$ & $\mathbb{Z}^{\times 2}$ & $0$ & 24, PQC & \begin{tabular}{c}$\epsilon_c=1$, $\eta_c=1$, $\epsilon_{pq}=-1$, $\epsilon_{pc}=1$, $\epsilon_{qc}=1$\\\hline$\epsilon_c=-1$, $\eta_c=1$, $\epsilon_{pq}=-1$, $\epsilon_{pc}=1$, $\epsilon_{qc}=-1$\\\hline$\epsilon_c=1$, $\eta_c=1$, $\epsilon_{pq}=-1$, $\epsilon_{pc}=1$, $\epsilon_{qc}=-1$\\\hline$\epsilon_c=-1$, $\eta_c=1$, $\epsilon_{pq}=-1$, $\epsilon_{pc}=1$, $\epsilon_{qc}=1$\end{tabular} &  $\mathcal{C}_0$ & $\mathbb{Z}$ & $0$ & $\mathbb{Z}$ & $0$\\\hline6, C & $\epsilon_c=1$, $\eta_c=1$ &  $\mathcal{R}_7$ & $0$ & $0$ & $0$ & $\mathbb{Z}$ & 25, PQC & \begin{tabular}{c}$\epsilon_c=1$, $\eta_c=-1$, $\epsilon_{pq}=-1$, $\epsilon_{pc}=1$, $\epsilon_{qc}=1$\\\hline$\epsilon_c=-1$, $\eta_c=-1$, $\epsilon_{pq}=-1$, $\epsilon_{pc}=1$, $\epsilon_{qc}=-1$\\\hline$\epsilon_c=1$, $\eta_c=-1$, $\epsilon_{pq}=-1$, $\epsilon_{pc}=1$, $\epsilon_{qc}=-1$\\\hline$\epsilon_c=-1$, $\eta_c=-1$, $\epsilon_{pq}=-1$, $\epsilon_{pc}=1$, $\epsilon_{qc}=1$\end{tabular} &  $\mathcal{C}_0$ & $\mathbb{Z}$ & $0$ & $\mathbb{Z}$ & $0$\\\hline7, C & $\epsilon_c=1$, $\eta_c=-1$ &  $\mathcal{R}_3$ & $0$ & $\mathbb{Z}_2$ & $\mathbb{Z}_2$ & $\mathbb{Z}$ & 26, PQC & \begin{tabular}{c}$\epsilon_c=1$, $\eta_c=1$, $\epsilon_{pq}=1$, $\epsilon_{pc}=-1$, $\epsilon_{qc}=1$\\\hline$\epsilon_c=-1$, $\eta_c=-1$, $\epsilon_{pq}=1$, $\epsilon_{pc}=-1$, $\epsilon_{qc}=1$\\\hline$\epsilon_c=1$, $\eta_c=1$, $\epsilon_{pq}=1$, $\epsilon_{pc}=-1$, $\epsilon_{qc}=-1$\\\hline$\epsilon_c=-1$, $\eta_c=-1$, $\epsilon_{pq}=1$, $\epsilon_{pc}=-1$, $\epsilon_{qc}=-1$\end{tabular} &  $\mathcal{R}_7$ & $0$ & $0$ & $0$ & $\mathbb{Z}$\\\hline8, C & $\epsilon_c=-1$, $\eta_c=1$ &  $\mathcal{R}_3$ & $0$ & $\mathbb{Z}_2$ & $\mathbb{Z}_2$ & $\mathbb{Z}$ & 27, PQC & \begin{tabular}{c}$\epsilon_c=1$, $\eta_c=-1$, $\epsilon_{pq}=1$, $\epsilon_{pc}=-1$, $\epsilon_{qc}=1$\\\hline$\epsilon_c=-1$, $\eta_c=1$, $\epsilon_{pq}=1$, $\epsilon_{pc}=-1$, $\epsilon_{qc}=1$\\\hline$\epsilon_c=1$, $\eta_c=-1$, $\epsilon_{pq}=1$, $\epsilon_{pc}=-1$, $\epsilon_{qc}=-1$\\\hline$\epsilon_c=-1$, $\eta_c=1$, $\epsilon_{pq}=1$, $\epsilon_{pc}=-1$, $\epsilon_{qc}=-1$\end{tabular} &  $\mathcal{R}_3$ & $0$ & $\mathbb{Z}_2$ & $\mathbb{Z}_2$ & $\mathbb{Z}$\\\hline9, C & $\epsilon_c=-1$, $\eta_c=-1$ &  $\mathcal{R}_7$ & $0$ & $0$ & $0$ & $\mathbb{Z}$ & 28, PQC & \begin{tabular}{c}$\epsilon_c=1$, $\eta_c=1$, $\epsilon_{pq}=-1$, $\epsilon_{pc}=-1$, $\epsilon_{qc}=1$\\\hline$\epsilon_c=-1$, $\eta_c=-1$, $\epsilon_{pq}=-1$, $\epsilon_{pc}=-1$, $\epsilon_{qc}=-1$\end{tabular} &  $\mathcal{R}_0^{\times 2}$ & $\mathbb{Z}^{\times 2}$ & $0$ & $0$ & $0$\\\hline10, PC & \begin{tabular}{c}$\epsilon_c=1$, $\eta_c=1$, $\epsilon_{pc}=1$\\\hline$\epsilon_c=-1$, $\eta_c=1$, $\epsilon_{pc}=1$\end{tabular} &  $\mathcal{C}_1$ & $0$ & $\mathbb{Z}$ & $0$ & $\mathbb{Z}$ & 29, PQC & \begin{tabular}{c}$\epsilon_c=1$, $\eta_c=-1$, $\epsilon_{pq}=-1$, $\epsilon_{pc}=-1$, $\epsilon_{qc}=1$\\\hline$\epsilon_c=-1$, $\eta_c=1$, $\epsilon_{pq}=-1$, $\epsilon_{pc}=-1$, $\epsilon_{qc}=-1$\end{tabular} &  $\mathcal{R}_4^{\times 2}$ & $\mathbb{Z}^{\times 2}$ & $0$ & $\mathbb{Z}_2^{\times 2}$ & $\mathbb{Z}_2^{\times 2}$\\\hline11, PC & \begin{tabular}{c}$\epsilon_c=1$, $\eta_c=-1$, $\epsilon_{pc}=1$\\\hline$\epsilon_c=-1$, $\eta_c=-1$, $\epsilon_{pc}=1$\end{tabular} &  $\mathcal{C}_1$ & $0$ & $\mathbb{Z}$ & $0$ & $\mathbb{Z}$ & 30, PQC & \begin{tabular}{c}$\epsilon_c=-1$, $\eta_c=1$, $\epsilon_{pq}=-1$, $\epsilon_{pc}=-1$, $\epsilon_{qc}=1$\\\hline$\epsilon_c=1$, $\eta_c=-1$, $\epsilon_{pq}=-1$, $\epsilon_{pc}=-1$, $\epsilon_{qc}=-1$\end{tabular} &  $\mathcal{R}_2^{\times 2}$ & $\mathbb{Z}_2^{\times 2}$ & $\mathbb{Z}_2^{\times 2}$ & $\mathbb{Z}^{\times 2}$ & $0$\\\hline12, PC & \begin{tabular}{c}$\epsilon_c=1$, $\eta_c=1$, $\epsilon_{pc}=-1$\\\hline$\epsilon_c=-1$, $\eta_c=-1$, $\epsilon_{pc}=-1$\end{tabular} &  $\mathcal{R}_7^{\times 2}$ & $0$ & $0$ & $0$ & $\mathbb{Z}^{\times 2}$ & 31, PQC & \begin{tabular}{c}$\epsilon_c=-1$, $\eta_c=-1$, $\epsilon_{pq}=-1$, $\epsilon_{pc}=-1$, $\epsilon_{qc}=1$\\\hline$\epsilon_c=1$, $\eta_c=1$, $\epsilon_{pq}=-1$, $\epsilon_{pc}=-1$, $\epsilon_{qc}=-1$\end{tabular} &  $\mathcal{R}_6^{\times 2}$ & $0$ & $0$ & $\mathbb{Z}^{\times 2}$ & $0$\\\hline13, PC & \begin{tabular}{c}$\epsilon_c=1$, $\eta_c=-1$, $\epsilon_{pc}=-1$\\\hline$\epsilon_c=-1$, $\eta_c=1$, $\epsilon_{pc}=-1$\end{tabular} &  $\mathcal{R}_3^{\times 2}$ & $0$ & $\mathbb{Z}_2^{\times 2}$ & $\mathbb{Z}_2^{\times 2}$ & $\mathbb{Z}^{\times 2}$ & 32, PQC & \begin{tabular}{c}$\epsilon_c=1$, $\eta_c=1$, $\epsilon_{pq}=1$, $\epsilon_{pc}=1$, $\epsilon_{qc}=-1$\\\hline$\epsilon_c=-1$, $\eta_c=1$, $\epsilon_{pq}=1$, $\epsilon_{pc}=1$, $\epsilon_{qc}=-1$\end{tabular} &  $\mathcal{R}_5$ & $0$ & $\mathbb{Z}$ & $0$ & $\mathbb{Z}_2$\\\hline14, QC & $\epsilon_c=1$, $\eta_c=1$, $\epsilon_{qc}=1$ &  $\mathcal{R}_0$ & $\mathbb{Z}$ & $0$ & $0$ & $0$ & 33, PQC & \begin{tabular}{c}$\epsilon_c=1$, $\eta_c=-1$, $\epsilon_{pq}=1$, $\epsilon_{pc}=1$, $\epsilon_{qc}=-1$\\\hline$\epsilon_c=-1$, $\eta_c=-1$, $\epsilon_{pq}=1$, $\epsilon_{pc}=1$, $\epsilon_{qc}=-1$\end{tabular} &  $\mathcal{R}_1$ & $\mathbb{Z}_2$ & $\mathbb{Z}$ & $0$ & $0$\\\hline15, QC & $\epsilon_c=1$, $\eta_c=-1$, $\epsilon_{qc}=1$ &  $\mathcal{R}_4$ & $\mathbb{Z}$ & $0$ & $\mathbb{Z}_2$ & $\mathbb{Z}_2$ & 34, K & $\eta_k=1$ &  $\mathcal{R}_1$ & $\mathbb{Z}_2$ & $\mathbb{Z}$ & $0$ & $0$\\\hline16, QC & $\epsilon_c=-1$, $\eta_c=1$, $\epsilon_{qc}=1$ &  $\mathcal{R}_2$ & $\mathbb{Z}_2$ & $\mathbb{Z}_2$ & $\mathbb{Z}$ & $0$ & 35, K & $\eta_k=-1$ &  $\mathcal{R}_5$ & $0$ & $\mathbb{Z}$ & $0$ & $\mathbb{Z}_2$\\\hline17, QC & $\epsilon_c=-1$, $\eta_c=-1$, $\epsilon_{qc}=1$ &  $\mathcal{R}_6$ & $0$ & $0$ & $\mathbb{Z}$ & $0$ & 36, PK & $\eta_k=1$, $\epsilon_{pk}=1$ &  $\mathcal{R}_1^{\times 2}$ & $\mathbb{Z}_2^{\times 2}$ & $\mathbb{Z}^{\times 2}$ & $0$ & $0$\\\hline18, QC & $\epsilon_c=1$, $\eta_c=1$, $\epsilon_{qc}=-1$ &  $\mathcal{R}_6$ & $0$ & $0$ & $\mathbb{Z}$ & $0$ & 37, PK & $\eta_k=-1$, $\epsilon_{pk}=1$ &  $\mathcal{R}_5^{\times 2}$ & $0$ & $\mathbb{Z}^{\times 2}$ & $0$ & $\mathbb{Z}_2^{\times 2}$\\\hline19, QC & $\epsilon_c=1$, $\eta_c=-1$, $\epsilon_{qc}=-1$ &  $\mathcal{R}_2$ & $\mathbb{Z}_2$ & $\mathbb{Z}_2$ & $\mathbb{Z}$ & $0$ & 38, PK & \begin{tabular}{c}$\eta_k=1$, $\epsilon_{pk}=-1$\\\hline$\eta_k=-1$, $\epsilon_{pk}=-1$\end{tabular} &  $\mathcal{C}_1$ & $0$ & $\mathbb{Z}$ & $0$ & $\mathbb{Z}$\\\hline\end{tabular}
}
\caption{\label{ref:tab0} Periodic table for symmetry-protected topological phases protected by non-Hermitian  symmetries belonging to the Bernard-LeClair classes. For each symmetry class, we summarize the symmetries involved, the symmetry generator relations, the classifying space for a 0D system (dimensional shift proceeds the same way as the conventional periodic table), as well as the topological classification in 0-3 dimensions.}
\end{table*}

\subsection{Unitary, Commuting Symmetries}
\label{sec:commutingsym}
The combination of several symmetries may give rise to a unitary symmetry, denoted as $M$, which commutes with the Hamiltonian. For the specific Hamiltonians we are considering, there are only two possibilities: $M=\bar{Q}$ or $M=\Sigma\bar{P}$, which both satisfy $M^2=+1$. 

Depending on the commutation relations with other symmetries, this may either result in multiple symmetry sectors, in which we should consider the problem separately in each symmetry sector, or generate a complex unit $J$ that transforms a real Clifford algebra into a complex one. When both unitary symmetries exist but do not commute with each other, the results could also depend on the order of diagonalizing into the symmetry sectors. Here, we choose to always first go into the symmetry sectors of the symmetry $\bar{Q}$ first, and then inspect whether a unitary, commuting symmetry $\Sigma \bar{P}$ still exists within each symmetry sector of $\bar{Q}$. 

In the following, we explain in detail our procedure for dealing with these possibilities.

\begin{enumerate}
    \item \textit{All remaining symmetry generators commute with $M$.} In this case, we can directly separate the problem into two symmetry sectors, with eigenvalues of $M$ being $\pm1$, while retaining all other symmetries. This will usually result in a doubling of the classification; however as shown in the next section, in the case of a type $Q$ symmetry, one can show that this doubling is in fact not physically meaningful, as the two sectors can be directly related by a symmetry operation.
    \item \textit{One unitary symmetry generator anti-commutes with $M$, while the rest commute with $M$.} In this case, we can still go into the $\pm 1$ eigenspace of $M$, except that now we should drop the symmetry generator that anti-commutes with $M$, since it is no longer a symmetry in each subspace.
    \item \textit{One anti-unitary symmetry generator anti-commutes with $M$, while the rest commute with $M$.} As discussed above, without loss of generality, we only need to consider at most one anti-unitary symmetry generator at a time. Therefore, the generator $JM$, where $J$ is the imaginary unit, actually commutes with all other symmetry generators. However, since $(JM)^2=-1$, $JM$ effectively acts as a complex unit ``$i$'' and reduces a real Clifford algebra into a complex one. Thus, we can obtain the classification using all symmetry generators other than $M$, and then reduce any real classes into complex classes by the identity \cite{morimoto2013topological} $Cl_{p,q}\otimes_\mathbb{R}Cl_{1,0}\simeq Cl_{p+q}$.
    \item \textit{Multiple symmetry generators anti-commute with $M$.} In this case, we take the first unitary symmetry that anti-commutes with $M$ and multiply it onto the remaining ones that anti-commute with $M$. The resulting generators will now all commute with $M$, so we have reduced the problem to that of case 2, where we simply discard the first non-commuting generator when we go into each symmetry sector.
\end{enumerate}
 
This allows us to take care of any unitary, commuting symmetries that may have exist as a consequence of the symmetries we specified.

\subsection{Relation between Type $Q$ Symmetry Sectors}
\label{sec:typeqsym}
For the unitary, commuting symmetry generated by a type $Q$ symmetry, it turns out that the two symmetry sectors will always be directly related by a transformation, and hence the sectors are in fact not independent. To see this, consider the doubled Hamiltonian $H_u$ defined above with symmetry $\bar{Q}$. The symmetry can be easily block-diagonalized by
\begin{align}
U\bar{Q}U^\dagger=\begin{pmatrix}
q & 0 \\ 0 & -q
\end{pmatrix},\quad 
U=\frac{1}{\sqrt{2}}\begin{pmatrix}
\mathbb{I} & \mathbb{I}\\\mathbb{I} & -\mathbb{I}
\end{pmatrix}.
\end{align}
Correspondingly, the doubled Hamiltonian is transformed into
\begin{align}\label{eqn:H_doubled}
UH_uU^\dagger=\begin{pmatrix}
\frac{u+u^\dagger}{2} & \frac{u^\dagger-u}{2}\\ \frac{u-u^\dagger}{2} & -\frac{u+u^\dagger}{2}
\end{pmatrix}.
\end{align}

We can diagonalize $q$ with a unitary matrix, and this will preserve the structure in Eq.~\ref{eqn:H_doubled}; moreover, since $q^2=I$,  we will only have eigenvalues of $\pm1$. Therefore, without loss of generality, we assume that $q$ is diagonal with the form $q=\textrm{diag}(\mathbb{I}_1,-\mathbb{I}_2)$, where $\mathbb{I}_1$ and $\mathbb{I}_2$ are identity matrices. We can move to the basis where $q$ is diagonalized and ordered in such a way. Since $qu^\dagger q^\dagger=u$, this implies that if we write $u$ in this basis, we must have
\begin{align}
u=\begin{pmatrix}
a & b \\  c & d
\end{pmatrix}\Rightarrow
u^\dagger=q^\dagger uq=\begin{pmatrix}
a & -b \\ -c &  d
\end{pmatrix},
\end{align}
such that the doubled Hamiltonian now takes the form
\begin{align}
UH_uU^\dagger=\begin{pmatrix}
a & 0  & 0 & -b\\ 0 & d & -c & 0\\ 0 & b & -a & 0\\c &  0 & 0 & -d\\
\end{pmatrix},
\end{align}
where the central block corresponds to eigenvalue $-1$,  while the corner blocks correspond to eigenvalue $+1$. On the other hand, we know that
\begin{align}
\begin{pmatrix}
0  & \mathbb{I}\\\mathbb{I}  & 0
\end{pmatrix}
\begin{pmatrix}
a & -b \\ c  & -d
\end{pmatrix}
\begin{pmatrix}
0  & \mathbb{I}\\\mathbb{I}  & 0
\end{pmatrix}
=-\begin{pmatrix}
d  & -c\\b  &  -a
\end{pmatrix}.
\end{align}
Therefore, the two symmetry sectors of the Hamiltonian are directly related by a symmetry transformation, up to a sign change. This implies that the topological properties of the two classes will always be locked to be the same. 

Thus, we only need to consider one of the symmetry sectors of the doubled $\bar{Q}$ symmetry. This also makes the resulting classification consistent with the well-known Hermitian AZ classes; otherwise, the classification obtained with our method would be doubled compared to the conventional AZ classes.

Here, we would like to remark that even though the inherent chiral symmetry $\Sigma$ does not always enter as a Clifford algebra generator in these cases (e.g. rows 14-21 in TABLE.~\ref{ref:tab1}), it still plays the important role of restricting the form of the Hamiltonian in Eq.~\ref{eqn:H_doubled}, which is necessary for showing that the two sectors are directly related to each other.

Note also that if we repeat the same analysis for the unitary, commuting symmetry $\Sigma \bar{P}$, we find that the doubled Hamiltonian is constrained to have the form
\begin{align}
H_u=\begin{pmatrix}
0  &  0 & 0 & b\\0 & 0 & c & 0\\0 & c^\dagger & 0 & 0\\b^\dagger & 0 & 0 & 0
\end{pmatrix},
\end{align}
such that the two symmetry sectors, formed by the central 2-by-2 and the corner 2-by-2, are independent from each other. Therefore, we can separately define topological invariants in each symmetry sector, and the classification is indeed doubled.

\subsection{Topological Classification}
With the above procedure, we have taken care of any potential unitary, commuting symmetries in the doubled Hamiltonian. Working in the remaining symmetry sector, the problem has been reduced to that of classifying a Hermitian Hamiltonian $H_u$ with a set of known symmetries of the usual form, which allows us to follow standard procedures of constructing the Clifford algebra extension problem and determining the topological classification, as reviewed in Sec.~\ref{sec:classifyherm}.

In our case, at this point of the analysis, there is at most one anti-unitary and one unitary symmetry generator that must anti-commute with the Hamiltonian. The only remaining step then is to multiply $J$ to any unitary symmetry generators or Hamiltonian terms that commute with the anti-unitary symmetry generator $e_a$, so that all the generators anti-commute with each other. We then also include the anti-unitary symmetry generator $Je_a$, which will automatically satisfy the desired anti-commutation relations and is independent from $e_a$ due to the presence of the imaginary unit.

Different topological phases are obtained from distinct ways to gap out the massless Dirac Hamiltonian by $H_u=\sum_{i=1}^d k_i\gamma_i+e_0$, which correspond to inequivalent ways to add the generator $e_0=m$ or $e_0=Jm$ to the Clifford algebra formed by $\{s,\gamma\}$, where $s$ are the set of symmetries, $\gamma$ are the set of momentum coefficient matrices, $m$ is the mass term and $J$ the imaginary unit. The set of representations for $e_0$ form the classifying space, denoted as $\mathcal{C}_k$ or $\mathcal{R}_k$ for complex and real classes. 

For complex classes, where the Clifford algebra extension problem is from $Cl_n\rightarrow Cl_{n+1}$, the classifying space is $\mathcal{C}_n$; for real classes, if the mass generator takes the form $m$, then $e_0^2=m^2=1$, and we have the extension problem $Cl_{p,q}\rightarrow Cl_{p,q+1}$, for which the classifying space is $\mathcal{R}_{q-p}$; if the mass generator takes the form $Jm$, then $e_0^2=(Jm)^2=-1$, and we have the extension problem $Cl_{p,q}\rightarrow Cl_{p+1,q}$, for which the classifying space is $\mathcal{R}_{p+2-q}$. The topological classification is then obtained from counting the connected components of the classifying space.

At this point, we would like to note that since momentum is reversed under time-reversal, $\gamma_i$ and $m$ will always have opposite commutation behavior with an anti-unitary symmetry generator $e_a$, and thus they must appear either as $\{J\gamma,m\}$ or $\{\gamma,Jm\}$ in the Clifford algebra. From our preceding discussion of the classifying space, this shows that as we increase the dimension of momentum space, the subscript of the classifying space decreases, which is the conventional shift direction, similar to Hermitian AZ classes.

Following through the complete procedure described above, we arrive at the classification table in TABLE.~\ref{ref:tab0}, where we enumerated each symmetry class with its characterization by the commutation relations and squaring properties of the symmetry generators. More details are presented in TABLE.~\ref{ref:tab1}, \ref{ref:tab2} in the appendix, where we have also provided the Clifford algebra generators, any commuting unitary symmetries\footnote{A subscript of the Clifford algebra bracket denotes that the generator commutes with all other symmetries and give rise to symmetry sectors; a separate bracket indicates that the generator serves as an imaginary unit}, as well as the classifying space and topological classification in physical dimensions 0 to 3. As discussed in Sec.~\ref{sec:equivrel}, some of the symmetry relations are equivalent to each other, and hence belong to the same topological class. For such cases, we include all equivalent representations in the table.

Moreover, we label the symmetry classes that have been discussed in the literature previously, such as the standard AZ classes \cite{altland1997nonstandard,ryu2010topological} as well as the non-Hermitian classes described in Ref.~\onlinecite{gong2018topological}. As expected, our classification results agree with prior ones where they overlap, but also provides a broad range of new symmetry classes with nontrivial topological properties.

\subsection{Obtaining Explicit Topological Invariants for Systems with Type $Q$ Symmetry}
\label{sec:altq}
In this section, we present an alternative approach to obtain the classification for symmetry combinations that contain a type $Q$ symmetry, without resorting to the doubled Hamiltonian.

Consider a non-Hermitian Hamiltonian $h$ that possesses a type $Q$ symmetry, and possibly an additional type $P$ symmetry and type $C$ symmetry. We can rewrite the type $Q$ symmetry condition as
\begin{align}
h=qh^\dagger q^\dagger \Rightarrow hq=(hq)^\dagger,
\end{align}
where we have used the fact that $q^2=\mathbb{I}$. Thus, we can directly construct a Hermitian Hamiltonian for these symmetry classes, without resorting to the preceding procedure of doubling the Hamiltonian.

The other symmetries can be correspondingly adapted for $hq$. If $h$ possesses a type $C$ symmetry, then $hq$ will also possess a type $C$ symmetry, with the symmetry now being implemented by $cq^*$. The parameters for the type $C$ symmetry now become
\begin{align}
\epsilon_c'=\epsilon_c\epsilon_{cq},\quad \eta_c'=\eta_c\epsilon_{cq},\quad \epsilon_{cp}'=\epsilon_{cp}\epsilon_{pq}.
\end{align}
If $h$ possesses a type $P$ symmetry, then we shall find that
\begin{align}
hq=-\epsilon_{qp}p(hq)p^\dagger.
\end{align}
Thus, if $\epsilon_{qp}=1$, then the constructed Hermitian Hamiltonian also possesses a chiral symmetry; otherwise, $p$ is a unitary symmetry that commutes with the Hamiltonian. Similar to the discussion in Sec.~\ref{sec:commutingsym}, if $p$ commutes with the anti-unitary symmetry $cq^*$, then we can treat it as a unitary, commuting symmetry and perform the analysis within each symmetry sector; otherwise, we multiply $p$ by the imaginary unit $J$ such that it commutes with $cq^*$, in which case $Jp$ acts as a complex number that reduces real classes into complex classes.

Since $q$ is invertible, the singular points where $h$ touches the base point $E=0$ will coincide with those where $hq$ touches the base point, and correspondingly, the topological classification will be the same between the two Hamiltonians. Thus, we can apply the standard Hermitian classification schemes to $hq$ to extract the topological properties of $h$. We have confirmed that this approach gives consistent results as the preceding classification table. In addition, this explicit connection also allows us to directly extract the topological invariants of non-Hermitian systems with type $Q$ symmetry, by simply mapping the Hamiltonian onto its Hermitian counterpart $hq$, and using well-known interpretations of topological invariants for Hermitian SPTs. However, such an approach only works for Hamiltonians that possess a type $Q$ symmetry, and the remaining symmetry classes still require the more general approach discussed in the preceding sections.

We note that more generally, there is another way one can obtain explicit non-Hermitian topological invariants from known Hermitian invariants. Due to the presence of the inherent chiral symmetry $\Sigma$ in the doubled Hamiltonian, the non-Hermitian Hamiltonian is precisely the block off-diagonal projector $q(k)$ defined in Ref.~\onlinecite{ryu2010topological}, and thus methods to obtain topological invariants for Hermitian Hamiltonians based on generalized winding numbers of the block off-diagonal projector can be directly applied to the non-Hermitian case.

\section{Examples of Classification and Construction of Explicit Invariants}
\label{sec:newclass}
We now discuss the preceding classification results in more detail by analyzing some specific examples, in order to elucidate the nature of the invariants and describe some useful techniques to directly obtain topological invariants. More specifically, we provide a detailed analysis of 0D pseudo-Hermitian systems (class 3), which possess a $\mathbb{Z}$ classification, and 1D systems with a type $C$ symmetry (class 7), which possess a $\mathbb{Z}_2$ classification, in order to illustrate some nontrivial examples beyond what has been analyzed before. Finally, we also comment on invariants for 2D systems, and describe how they can be understood from winding numbers associated with block off-diagonal projectors in chiral Hermitian Hamiltonians. While we illustrate the results for specific examples, many of the techniques can be readily generalized to other symmetry classes and dimensions.

\subsection{Zero-Dimensional Systems with $\mathbb{Z}$ Classification}
In previous non-Hermitian SPT classification attempts \cite{gong2018topological}, all $0$D systems belonged to the $\mathbb{Z}_2$ or trivial classification, and the $\mathbb{Z}_2$ topological invariant was interpreted as the parity of the number of negative real eigenvalues. However, our analysis shows that there exist several $0$D non-Hermitian classes which are classified by a $\mathbb{Z}$ invariant. Indeed, a general Hermitian system without any additional symmetries is expected to possess a $\mathbb{Z}$ classification. However, we also find other classes, with a somewhat different topological interpretation compared to the Hermitian case.

First, let us consider a non-Hermitian Hamiltonian possessing a type $Q$ symmetry (class 3 in TABLE.~\ref{ref:tab0}). Without loss of generality, we have two possibilities for $q$, being either $I$ or $\sigma_3$ in an appropriate basis ($\sigma_{1,2,3}$ are the Pauli matrices, and below we write $\sigma_{ij}=\sigma_i\otimes \sigma_j$ etc.). The former one implies that the system is simply Hermitian $h = h^\dagger$, and the latter one implies that the system is pseudo-Hermitian, $h = \sigma_3 h^\dagger \sigma_3$. In both cases, we obtain a $\mathbb{Z}$ classification (see TABLE.~\ref{ref:tab0}); more precisely, we will show that an $N$-state system has $N+1$ distinct topological classes of Hamiltonians.

For the former case of a Hermitian matrix, the topological invariant is known to simply count the number of eigenvalues below zero. The latter case however requires more careful analysis; consider a two-band model, where we expect three different classes as in the Hermitian case. The pseudo-Hermitian model is constrained to have the following form
\begin{equation}
    h = a I + b i \sigma_1 + c i \sigma_2 + d \sigma_3, \quad a,b,c,d \in \mathbb{R},
    \label{eq:0DZ}
\end{equation}
where the eigenvalues are $a \pm \sqrt{d^2 - b^2 - c^2}$. This implies that the spectra consists of either two real eigenvalues, or a complex-conjugate pair. Crucially however, the emergence of a complex-conjugate pair of eigenvalues from two real eigenvalues requires the two real eigenvalues to coalesce first.

\begin{figure} [t]
 \includegraphics[width=0.48\textwidth]{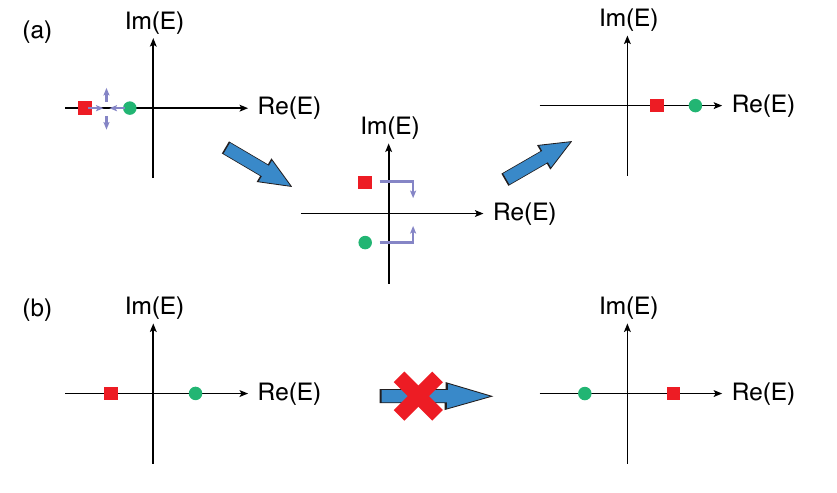}
 \vspace{-4pt}
\caption{\label{fig:0D_Z} $\mathbb{Z}$ classification of 0D topological classes with a type $Q$ symmetry $\sigma_z$ (class 3), illustrated for the case of a two-band model Eq.~\ref{eq:0DZ}. (a) Two eigenvalues on the negative real axis can combine, split into the complex plane, and move over to the positive real axis. (b) In contrast, an eigenvalue on the positive axis cannot swap places with an eigenvalue on the negative axis, because there are no eigenvalues on the same side to combine and move over in the complex plane.}
\end{figure}

For this eigenvalue spectrum, unlike the Hermitian case where $\mathbb{I} \nsim -\mathbb{I}$, we now have $\mathbb{I} \sim -\mathbb{I}$, since these two matrices can be connected by taking the following path:
\begin{equation}
    (a,b): (1,0) \mapsto (1,1) \mapsto (-1,1) \mapsto (-1,0),
\end{equation}
with $c=d=0$. In terms of eigenvalues, this corresponds to the path shown in Fig.~\ref{fig:0D_Z}(a). However, one can see that $h = \sigma_3$ and $h'= -\sigma_3$ are not connected, since the two eigenvalues of each matrix lie on the positive and negative real axis, and do not have partners on the same side to combine and split into the complex plane (see Fig.~\ref{fig:0D_Z}(b)). Therefore, in order for the two eigenvalues to switch places, they must pass through the origin, and must thus go through a topological phase transition. For more than two bands, there is a corresponding generalization of this intuition, where unlike the $\mathbb{Z}_2$ case \cite{gong2018topological}, only certain eigenvalues originating from different chiral symmetry sectors are allowed to combine and split into the complex plane, and move from the positive real axis to the negative real axis.

The different topological classes can also be seen more directly by observing that for a generic type $Q$ symmetry $q$, the symmetry condition implies that $hq$ is Hermitian, and since $q$ is an invertible matrix, there is a one-to-one correspondence between topological equivalence classes of $h$ and $hq$ (see Sec.~\ref{sec:altq}). Thus, the topological classes of $h$ are characterized by the number of negative  eigenvalues of the modified Hamiltonian $hq$. Studying the form of the doubled Hamiltonian as in Sec.~\ref{sec:typeqsym}, one obtains similar results.

\subsection{One-Dimensional Systems with $\mathbb{Z}_2$ Classification}
\label{sec:1D}
In 1D, previous classification results in Ref.~\onlinecite{gong2018topological} described the existence of a $\mathbb{Z}$ invariant, characterizing the winding number of the bands around the origin of the complex energy plane $E=0$. However, we find that our periodic table gives rise to more possibilities when generic non-Hermitian symmetries are considered. Here, we present a model with a $\mathbb{Z}_2$ invariant. This model does not possess a type $Q$ symmetry, and thus the nature of the invariant cannot be simply understood from the transformation described in Sec.~\ref{sec:altq}.

Consider symmetry class 7 in TABLE.~\ref{ref:tab0}, which possesses only a type $C$ symmetry, where $c h^T c^\dagger = h$ and $c c^*= -1$. We construct a Hamiltonian realizing this symmetry class by making use of the following complex representation of the Clifford algebra: 
\begin{center}
\begin{tabular}{c|c|c|c|c|c}
Gen. & $\Sigma$ & $\bar{C}$ & $J\bar{C}$ & $\gamma_x$ & $Jm$  \\
\hline
\, Rep. \, & \,\, $\sigma_{30}$ \,\,& \,\, $\sigma_{12} \mathcal{K}$ \,\,&\,\, $i\sigma_{12}\mathcal{K}$ \,\,&\,\, $\sigma_{21}$ \,\,&\,\, $i\sigma_{10}$ \,\,\\
\end{tabular}
\end{center}
where we have satisfied all constraints on the realization of generators, such that $\Sigma$ always takes the form of $\sigma_{3} \otimes I$, and $\bar{C}$ takes the form of $(\sigma_{1} \otimes M) {\cal K}$ (see Eq.~\ref{eq:doubsym}). Using the given representations of kinetic and mass terms, one can construct the following non-Hermitian Hamiltonian:
\begin{equation}\label{eq:1D_Z2_model}
    h(k_x) = i \sin k_x \sigma_{1} + (m - \cos k_x) \sigma_{0},
\end{equation}
where $m$ is a tuning parameter and the symmetry is implemented as $c = \sigma_{2}$. Here, in moving from a low-energy continuum description to a lattice model, we have chosen a momentum-dependent coefficient $m-\cos k_x$ for the mass term, so that the system is gapless only at $k_x = 0$ and $m=1$. If the mass term is chosen to be independent of $k_x$, then there will be two Dirac points at $k_x=0,\pi$ that will be gapped out together, corresponding to two topological transitions happening simultaneously.

\begin{figure} [t]
 \includegraphics[width=0.48\textwidth]{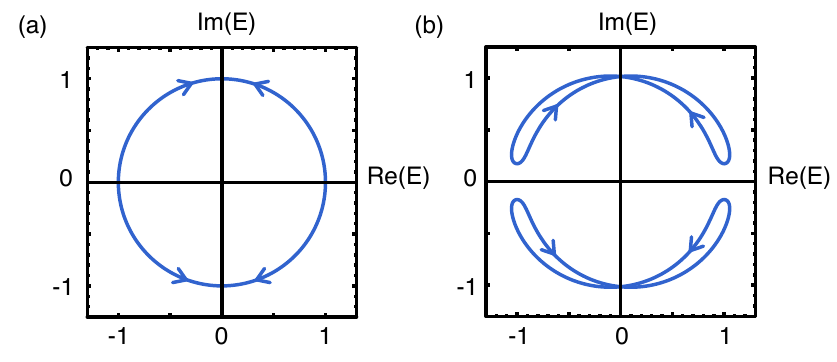}
 \vspace{-4pt}
\caption{\label{fig:1D_Z2} $\mathbb{Z}_2$ classification of 1D topological classes with type $C$ symmetry (class 7), for the case of two copies of the two-band model Eq.~\ref{eq:1D_Z2_model}. (a) Spectra for a single copy, with two degenerate bands. One band has winding number $+1$, and the other has winding number $-1$. (b) Coupling two copies of the model under symmetry-respecting perturbations. Here, bands split in such a way that all bands have zero winding number with respect to the origin.}
\end{figure}

There are two bands in the Hamiltonian Eq.~\ref{eq:1D_Z2_model}, with energies $E_\pm = m-\cos k_x \pm i \sin k_x$. The $E_+$ and $E_-$ branches have opposite winding-numbers around the origin, and they are related by the type $C$ symmetry as $E_+(-k_x) = E_-(k_x)$. The two different topological phases ($m>1$ or $m<1$) for this model can then be characterized by the parity of a winding number, defined as
\begin{equation}
    w' = \int_{-\pi}^{\pi} \frac{dk}{4\pi i}   \partial_{k_x} \qty( \arg E_+(k_x) - \arg E_-(k_x) ).
\end{equation}
where $(-1)^{w'} = -1$ for the topologically non-trivial phase and $(-1)^{w'} = 1$ for the trivial phase. This definition can be generalized into any model with $2n$ bands for this symmetry class, since the generalized Kramers degeneracy (Sec.~\ref{sec:typeckramers}) ensures that all bands come in pairs, with opposite chiralities. Unlike the previously-studied 1D non-Hermitian classes with a $\mathbb{Z}$ classification \cite{gong2018topological}, this symmetry class has a $\mathbb{Z}_2$ classification. Thus, we observe robust helical modes with $\Re(E)=0$ and $\Im(E) > 0$ in Fig.~\ref{fig:1D_Z2}(a), implying that the physical response of the system at late times depends on a pair of counter-propagating modes. We remark that this feature is very similar to the edge of a 2D topological insulator (TI)---class 15 (Hermitian class AII) in our table. Considering that $(i)$ the symmetry of class 7 gives rise to a non-Hermitian Kramers degeneracy and $(ii)$ classifications of class 7 in $d$-dimension and class 15 at $(d+1)$ dimension exactly match, one may conjecture that there exists a higher-level of correspondence between non-Hermitian and Hermitian classifications for these specific cases.

To demonstrate how the classification becomes $\mathbb{Z}_2$ instead of $\mathbb{Z}$, let us consider two copies of the aforementioned system, and analyze the effects of symmetry-preserving perturbations. When we have two copies of Eq.~\ref{eq:1D_Z2_model}, the symmetry action becomes $c = \sigma_{20}$. Under this symmetry, the following mass terms (coupling between two copies) are allowed: $\sigma_{00}, \sigma_{01},  \sigma_{12}, \sigma_{22}, \sigma_{32}, \sigma_{03}$. For simplicity, we consider 
\begin{equation}
    h'(k_x) = i \sin k_x \sigma_{10} + (m - \cos k_x ) \sigma_{00} + \delta (\sigma_{12} + 2 i \sigma_{32}),
\end{equation}
where the mass term $\sigma_{12} + 2 i \sigma_{32}$ is chosen to break degeneracies. At $m=\delta = 0$, we obtain Fig.~\ref{fig:1D_Z2}(a), where four bands are quadruply degenerate, and we have $\abs{w'} = 2$. However, at $m=0$ and $\delta = 0.1$, the spectra becomes that of Fig.~\ref{fig:1D_Z2}(b) where the four bands split into doubly degenerate bands with $\abs{w'} = 0$. Thus, this phase coincides with two copies of the topologically trivial phase ($m=2$ and $\delta = 0$), demonstrating that the classification is $\mathbb{Z}_2$, and the newly defined winding number $w'$ is defined modulo $2$.

\subsection{Two-Dimensional Systems with $\mathbb{Z}_2$ Classification}
\label{sec:2D}
Moving up another dimension, we provide an example in which the system has a nontrivial classification in 2D. Note that in previous non-Hermitian classifications \cite{gong2018topological}, all 2D classes were found to be trivial.

The simplest example is class 3, which possesses a type $Q$ symmetry. For a Hamiltonian with a type $Q$ symmetry, the invariant can be obtained similarly to the 0D example, in which we may transform the Hamiltonian by the type $Q$ symmetry to obtain the topological invariant of the non-Hermitian system from the corresponding Hermitian one.

In the absence of a type $Q$ symmetry, the first non-trivial 2D example is again given by class 7, with a type $C$ symmetry. To illustrate the topological classification for this symmetry class, we make use of the following complex representation  of the Clifford algebra
\begin{center}
\begin{tabular}{c|c|c|c|c|c|c}
Gen. & $\Sigma$ & $\bar{C}$ & $J\bar{C}$ & $\gamma_x$ & $\gamma_y$ & $Jm$  \\
\hline
\, Rep. \, & \,\, $\sigma_{30}$ \,\,& \,\, $\sigma_{12} \mathcal{K}$ \,\,&\,\, $i\sigma_{12}\mathcal{K}$ \,\,&\,\, $\sigma_{11}$ \,\,& \,\, $\sigma_{12}$ \,\,&\,\, $i\sigma_{20}$ \,\,\\
\end{tabular}
\end{center}
This gives rise to the following non-Hermitian Hamiltonian on a lattice:
\begin{align} \label{eq:2D_Ham}
    h(k_x,k_y) &= \sin  k_x\sigma_1+\sin k_y\sigma_2\nonumber\\&-i[m+C(\cos k_x+\cos k_y)],
\end{align}
with the corresponding doubled Hamiltonian given by
\begin{align}
H(k_x,k_y)&= \sin  k_x\sigma_{11}+\sin k_y\sigma_{12}\nonumber\\&+[m+C(\cos k_x+\cos k_y)]\sigma_{20}.
\end{align}

\begin{figure}[t]
    \centering
     \includegraphics[width=0.47\textwidth]{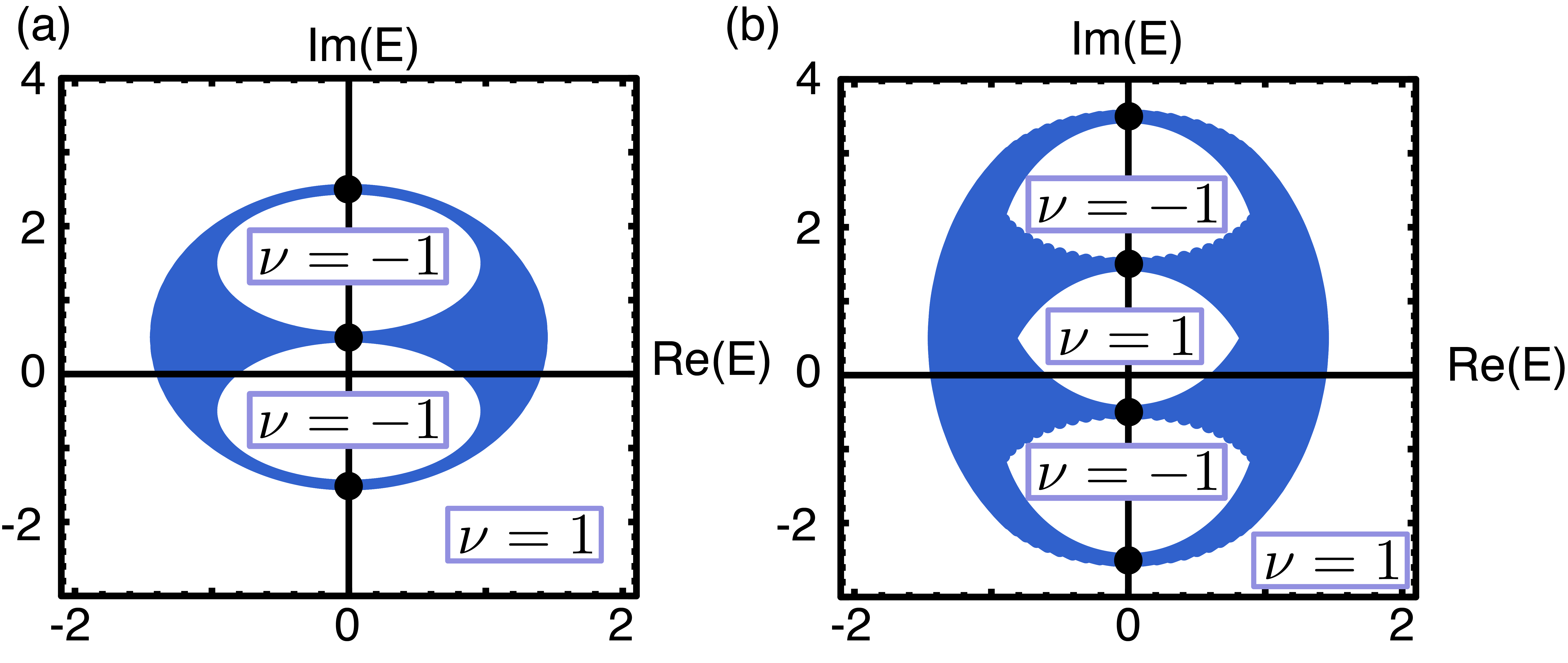}
    \caption{$\mathbb{Z}_2$ classification of 2D topological classes with a type $C$ symmetry (class 7). Since the Brillouin zone is now two-dimensional, the complex spectra is also a two-dimensional object. The system carries a different topological index $\nu$ depending on where the origin (base energy) is located at, as labeled. Black dots represent a Dirac cone in the real part of the spectrum, where topological phase transitions occur. (a) Spectra of Eq.~\ref{eq:2D_Ham}, where there is no anisotropy between $x$ and $y$ directions. Here, $m=0.5$ and $C=1$. (b) Spectra for the Hamiltonian Eq.~\ref{eq:2D_mass} with a modified mass term. Here, $m=0.5$, $C_x = 1$, and $C_y = 2$.
    }
    \label{fig:2D_Bands}
\end{figure}

Upon closer inspection, we find that the doubled Hamiltonian is identical to the 2D class DIII example described in Sec.~4.1.4 of Ref.~\onlinecite{ryu2010topological}. At the doubled level, the $\mathbb{Z}_2$-invariant can be expressed as the Kane-Mele invariant  \cite{kane2005z}, which can be reduced to the Pfaffian of the sewing matrix (characterizing the transformation of eigenstates under symmetry operations) at time-reversal invariant points \cite{kane2005z,ryu2010topological,bernevig2013topological}. Evaluating the Pfaffian, we find that the $\mathbb{Z}_2$-invariant is given by
\begin{eqnarray}
    \nu &=& \prod_{\vec{k} \in \textrm{TRI}}  \textrm{Pf}[w(\vec{k})] = \prod_{\vec{k} \in \textrm{TRI}}  \frac{m(\vec{k})}{\abs{m(\vec{k})}}  \nonumber \\
    &=& \textrm{sgn}(m+2C) \, \qty[\textrm{sgn}(m)]^2 \, \textrm{sgn}(m-2C),
\end{eqnarray}
where $m(\vec{k}) = m+C(\cos k_x+\cos k_y)$ and $w(\vec{k})$ is the sewing matrix defined in Ref.~\onlinecite{ryu2010topological}. Thus, the phase with $-2C < m < 2C$ is non-trivial while the phase with $m > 2C$ or $m < -2C$ is trivial. In Fig.~\ref{fig:2D_Bands}(a), we plot the dispersion of Eq.~\ref{eq:2D_Ham} for $m=0.5$, $C=1$. As one can see, there are two `holes' in the two-dimensional dispersion drawn in the complex energy plane. The system undergoes a topological phase transition as the origin moves into the holes, resulting in different topological classes depending on where the origin is located relative to the spectrum.

We can also consider a more generic case in which there is an anisotropy between the $x$ and $y$ directions. Here, the mass term is given by 
\begin{align}
    m(\vec{k}) &= m + C_x \cos (k_x) + C_y \cos(k_y),\label{eq:2D_mass}
\end{align}
where the expression for topological indices is modified accordingly. In this case, there are three holes in the complex spectra (see  Fig.~\ref{fig:2D_Bands}(b)), the spectra in Fig.~\ref{fig:2D_Bands}(a) being the special case where $C_x = C_y$, so that the middle hole disappears. Black dots in the figure represent Dirac points in the real part of the spectrum, and the topological transition in this specific model is also accompanied by a sign change of the imaginary energy of the Dirac point.

We can also directly obtain the invariant of the 2D system by making use of the correspondence between the non-Hermitian Hamiltonian and the block off-diagonal projector $q(k)$ \cite{ryu2010topological}, which characterizes one off-diagonal sub-block of a Hermitian Hamiltonian with chiral symmetry, in an appropriate basis. From TABLE.~\ref{ref:tab0}, we observe that the doubled Hamiltonian of symmetry class 7 has a classifying space $\mathcal{R}_3$, which is identical to a Hermitian system of symmetry class DIII. As discussed in detail in Ref.~\onlinecite{ryu2010topological}, for Hermitian class DIII, the two-dimensional $\mathbb{Z}_2$ invariant is the first descendant of a three-dimensional $\mathbb{Z}$ invariant. 

Thus, the topological invariant can be obtained by the following general procedure: First, identify a trivial phase corresponding to non-Hermitian Hamiltonian $h_0(k)$ (or equivalently, the block off-diagonal projector of the doubled Hamiltonian). Then, for a given Hamiltonian $h_1(k)$, its topological invariant relative to the trivial phase can be determined by constructing a continuous path $h(k,t)$, $t\in [0,\pi]$, such that $h(k,0)=h_0(k)$ and $h(k,\pi)=h_1(k)$ (The system must remain gapped throughout the trajectory). This path can then be expanded to a non-Hermitian Hamiltonian in one higher dimension $h(k,t)$, with $t\in [0,2\pi]$ now regarded as an additional momentum direction, such that the expanded Hamiltonian satisfies the symmetry constraint for class DIII $h(k,t)=-\bar{c} h^T(-k,2\pi-t) \bar{c}^\dagger$ where $\bar{c} = \sigma_1 \otimes c$. The winding number of this higher-dimensional  Hamiltonian can then be calculated
\begin{align}
    \nu_3[h(k,t)]=\int_{BZ^{d=3}}\frac{1}{24\pi^2}\textrm{tr}[(h^{-1}\text{d}h)^{3}],
\end{align}
and the topological invariant relative to Hamiltonian $h_0(k)$ will be
\begin{align}
    \nu=(-1)^{\nu_3[h(k,t)]},
\end{align}
since different path interpolations \cite{ryu2010topological} between $h_0(k)$ and $h_1(k)$ can be shown to be equivalent mod 2. Thus, using this procedure, it is possible to provide a topological invariant for a general 2D non-Hermitian Hamiltonian in class 7. As commented in the preceding sections, since the chiral symmetry is inherent to the doubling procedure, this technique is also applicable to many non-Hermitian Hamiltonians in other symmetry classes, at least for the ones which contain a chiral symmetry in the Clifford generator set and do not possess an imaginary unit that reduces real classes to complex classes (see third column in TABLE.~\ref{ref:tab1},~\ref{ref:tab2}).

\section{Summary and Discussion}
\label{sec:discussion}

In conclusion, we have classified non-Hermitian topological bands in arbitrary spatial dimensions, systematically accounting for the new types of symmetries that are unique to non-Hermitian systems. Making use of the Bernard-LeClair symmetry classes, we have found that these generic non-Hermitian symmetries give rise to a wide range of possibilities, where many symmetry classes that have thus far not been explored possess a nontrivial classification. The entire Hermitian AZ classification is also naturally incorporated as a special instance of our current classification scheme \cite{magnea2008random}, since Hermiticity is viewed in this picture as a type $Q$ symmetry, with the symmetry implemented as the identity matrix. 

In addition, by a direct mapping of the non-Hermitian Hamiltonian into a Hermitian one, we have elucidated how the nature of the topological invariant for symmetry classes with a type $Q$ symmetry can be understood from well-known examples in Hermitian systems. We have also provided new topological invariants for classes where such a transformation is not immediately available. In particular, we find that unlike previous work \cite{gong2018topological}, there do exist non-Hermitian models in 1D and 2D, with no direct transformation to Hermitian models, that possess a $\mathbb{Z}_2$ topological invariant.

Moreover, there seems to be an interesting correspondence between classifications of $d$-dimensional non-Hermitian systems and $d+1$-dimensional Hermitian systems, where the long time behavior of certain classes of non-Hermitian systems resemble the anomalous boundary of Hermitian systems. Considering that a non-Hermitian Hamiltonian can be regarded as an effective description of a Hermitian system where the bulk is integrated out, this may hint at the possibility of non-Hermitian classifications capturing anomalous physics in the same dimension.

Beyond its immediate importance for understanding the topological structure of non-Hermitian systems, the periodic table we have derived can also guide experimental design, where the symmetries that can give rise to novel topological effects are now known. Such results, in combination with further understanding of the unconventional bulk-boundary correspondence, may facilitate experimental studies in atomic and photonic systems. In addition, the formalism can be readily extended to include other types of symmetries such as crystalline symmetries \cite{morimoto2013topological,liu2018topological}, as well as studies of higher-order topological insulators in non-Hermitian systems \cite{benalcazar2017quantized,lee2018hybrid,liu2018second,edvardsson2018non,benalcazar2017electric,song2017,langbehn2017reflection,kunst2018lattice,schindler2018higher,khalaf2018higher,matsugatani2018connecting,imhof2017topolectrical,peterson2018quantized,serra-garcia2018observation,schindler2018higher2}. Finally, we would like to mention that the current classification makes use of one particular generalization of the Hermitian notion of a gapped phase, namely the prohibition of touching a base point \cite{gong2018topological}, but there may also be other generalizations, particularly ones related to spectral degeneracies such as exceptional points, where symmetry-protected nodal lines and nodal surfaces have been found in 2D and 3D \cite{budich2018symmetry,okugawa2018topological,zhou2018exceptional}.

\section*{Acknowledgments}
The authors thank D.~Borgnia, Z.~Gong, F.~Grusdt, W.~W.~Ho, S.~Lieu, N.~Rivera, R.-J.~Slager, S.~Vijay, A.~Vishwanath, Y.~You, M.~Zhang, B.~Zhen, L.~Zou for helpful discussions. H.~Z. thanks E.~Demler and F.~Grusdt for supervising the final presentation which partially inspired this work; H.~Z. and J.~Y.~L acknowledge the support of the Boulder summer school on condensed matter and material physics (quantum information), where this work was initiated.

Note added: after this work was near completion, we became aware of a related work by Kawabata, Shiozaki, Ueda, and Sato \cite{kawabata2018symmetry}. Although we initially counted the number of symmetry classes as 42, we revisited our calculations and removed a redundancy using Eq.~\ref{eq:p+q} after knowing the 38-fold symmetry classification in Ref.~\onlinecite{kawabata2018symmetry}.

\appendix
\section{Requirements on Square of Symmetry Operator}
\label{suppsec:symmetryclass}
As mentioned in the main text, the condition that the symmetries are involutions also imposes restrictions on the unitary symmetry implementation $u$. Acting twice with each of the symmetries, one obtains
\begin{align}
h&=-php^{-1}=p^2h\qty(p^2)^{-1},\\
h&=qh^\dagger q^{-1}=q\qty(qh^\dagger q^{-1})^\dagger q^{-1}=q^2 h\qty(q^2)^{-1},\\
h&=\epsilon_c ch^Tc^{-1}=c\qty(ch^T c^{-1})^T c^{-1}=\qty(cc^*)h\qty(cc^*)^{-1},\\
h&=kh^*k^{-1}=k\qty(k^*h^*k^T)k^{-1}=\qty(kk^*)h\qty(kk^*)^{-1}.
\end{align}
where the matrices $p$, $q$, $c$, $k$ are the unitary implementations of type $P$, $Q$, $C$, $K$ symmetries. Since the symmetry operations are required to be involutions and $h$ can be a generic Hamiltonian, by Schur's lemma we must have that $p^2=q^2=cc^*=kk^*=\lambda \mathbb{I}$. Moreover, since the matrices are unitary, $|\lambda|=1$; for $p$ and $q$, we can multiply a phase factor to make $\lambda=1$; for $c$ and $k$, we can show that $\lambda=\pm 1$ as follows:
\begin{align}
&u u^*=\lambda \mapsto u^* = \lambda u^\dagger \mapsto u = \lambda^* u^T \nonumber\\
&u u^*=\lambda \mapsto u = \lambda u^T \nonumber\\
&(\lambda - \lambda^*)u^T = 0
\end{align}
Thus $\lambda$ is real and being norm 1, must take on values $\lambda=\pm 1$. The same argument holds for the Hermitian anti-unitary symmetries, namely time-reversal symmetry and particle-hole symmetry.

\section{Details of Unitarization of Non-Hermitian Hamiltonian}
\label{suppsec:unitarization}
We now prove that the unitary matrix $u$ in the polar decomposition $h=uP$ of a non-Hermitian Hamiltonian has the same symmetry as $h$, for each of the four symmetry classes. As discussed in the main  text, we will make use of the following property: if $A$ and $B$ are positive-definite matrices, and $A^2=B^2$, then $A=B$, since a positive-definite matrix has a unique positive-definite square root.
\begin{enumerate}
\item Type P. We know that
\begin{align}
hu_p=-u_ph\Leftrightarrow uPu_p=-u_puP,
\end{align}
where we have written the unitary symmetry implementation as $u_p$ to emphasize its unitary nature. As above, $u$ is also unitary and $P$ Hermitian. Taking the complex conjugate of the preceding equation and multiplying it from the left on both sides, using the unitarity of $u$ and $u_p$ as well as the Hermiticity of $P$, we have
\begin{align}
(u_p^\dagger Pu_p)^2=P^2,
\end{align}
which by virtue of $P$ being positive-definite results in $u_p^\dagger Pu_p=P$ via the preceding property. Plugging this back in, we have
\begin{align}
uPu_p=uu_pP=-u_puP\Rightarrow uu_p=-u_pu,
\end{align}
where we have used the positive-definite property of $P$ in the last step. Thus $u$ satisfies the same symmetry as $h$.
\item Type Q. We follow the same recipe:
\begin{align}
&hu_q=u_qh^\dagger\Leftrightarrow uPu_q=u_qPu^\dagger\nonumber\\
&\Rightarrow (u_q^\dagger Pu_q)^2=(uPu^\dagger)^2
\Leftrightarrow u_q^\dagger Pu_q=uPu^\dagger \nonumber\\&\Rightarrow uu_q(uPu^\dagger)=u_qu^\dagger uPu^\dagger\Rightarrow uu_q=u_qu,
\end{align}
where in the last step we used the fact that $uPu^\dagger$ is positive-definite.
\item Type C.
\begin{align}
&hu_c=\epsilon_c u_ch^T \Leftrightarrow uPu_c=\epsilon_c u_cP^Tu^T \nonumber\\
&\Rightarrow (u_c^\dagger P u_c)^2=(u^* P^Tu^T)^2
\Leftrightarrow u_c^\dagger P u_c=u^* P^Tu^T \nonumber\\
&\Rightarrow uPu_c=uu_c(u^*P^Tu^T)=\epsilon_c u_cu^T(u^*P^Tu^T)\nonumber\\&\Rightarrow uu_c=\epsilon_c u_cu^T.
\end{align}
\item Type K.
\begin{align}
&hu_k=u_kh^*\Leftrightarrow uPu_k=u_ku^*P^*\nonumber\\
		&\Rightarrow (u_k^\dagger Pu_k)^2=(P^*)^2\Leftrightarrow u_k^\dagger Pu_k=P^*\nonumber\\
		&\Rightarrow uPu_k=uu_kP^*=u_ku^*P^*\nonumber\\
&\Rightarrow  uu_k=u_ku^*.
\end{align}
\end{enumerate}
This completes the proof.

\section{Constructing Concrete Examples of Classification}
In this appendix, we provide some more examples to illustrate the construction of non-Hermitian Hamiltonians from the Clifford algebra approach, and show that the resulting Hamiltonians have topological characteristics that are consistent with the results of the periodic table.

The general recipe is to start from the Clifford generators in TABLE.~\ref{ref:tab1} and \ref{ref:tab2}, and iterate through different combinations of Pauli operators to find a Clifford algebra realization that satisfies the commutation relations, squares, as well as the specified forms of the operators $\Sigma=\sigma_z\otimes \mathbb{I}$ and Eq.~\ref{eq:doubsym}. This discussion however is only valid within each symmetry sector of the doubled Hamiltonian, so to construct the full Hamiltonian we write down a model for each symmetry sector of the unitary commuting symmetry $\Sigma P$, and make use of the relationship between type $Q$ symmetry sectors (Sec.~\ref{sec:typeqsym}) to obtain the original non-Hermitian Hamiltonian with the desired non-Hermitian  symmetries.

As an example, let us consider the Hamiltonians for symmetry classes 14 and 18 in 0D. Class 14 corresponds to the Hermitian AI class, while class 18 differs due to the different commutation relation between $q$ and $c$. From TABLE.~\ref{ref:tab1}, we see that the different commutation relation drastically changes the classification, such that the $\mathbb{Z}$ classification for the Hermitian AI class now becomes trivial under class 18.

Using the Clifford algebra generators described in TABLE.~\ref{ref:tab1}, we find the simplest realization for class 14 has generators,
\begin{center}
\begin{tabular}{c|c|c|c}
Gen. & $\bar{C}$ & $J\bar{C}$ & $Jm$  \\
\hline
\, Rep. \, & \,\, $\sigma_{1} \mathcal{K}$ \,\,&\,\, $i\sigma_{1}\mathcal{K}$ \,\,&\,\, $i\sigma_0$\,\,\big| \, $i\sigma_1$\,\,\big| \, $i\sigma_2$\,\,\\
\end{tabular}
\end{center}
while for class 18,
\begin{center}
\begin{tabular}{c|c|c|c}
Gen. & $\Sigma\bar{C}$ & $J\Sigma\bar{C}$ & $m$  \\
\hline
\, Rep. \, & \,\, $\sigma_{2} \mathcal{K}$ \,\,&\,\, $i\sigma_{2}\mathcal{K}$ \,\,&\,\, $\sigma_1$\,\,\big| \, $\sigma_2$\,\,\big| \, $\sigma_3$\,\,\\
\end{tabular}
\end{center}

Since the class 14 Hamiltonian is Hermitian, we can easily right down the Hamiltonian
\begin{center}
\begin{tabular}{c|c|c|c}
Class & Hamiltonian & $C$ sym. & $Q$ sym.  \\
\hline
14 & $m_1\sigma_0+m_2\sigma_1+m_3\sigma_2$ & $\sigma_1$ & $\sigma_0$\\
\end{tabular}
\end{center}
where the mass terms $m_i$ are all real. The spectrum for the class 14 Hamiltonian is $E=m_1\pm \sqrt{m_2^2+m_3^2}$, where the system belongs to different phases depending on whether $m_1^2> m_2^2+m_3^2$ or $m_1^2< m_2^2+m_3^2$. Additionally, for $m_1^2>m_2^2+m_3^2$, the system belongs to two different phases depending on whether $m_1>0$ or $m_1<0$, since a continuous deformation between them necessarily touches $E=0$. Geometrically, the three phases are separated by the double cone $m_1=\pm \sqrt{m_2^2+m_3^2}$  in parameter space. Thus, there are three topological classes for this two-band model, consistent with the $\mathbb{Z}$ classification.

Now we consider the non-Hermitian Hamiltonian characterizing class 18. We choose the type $Q$ symmetry to be of the form $q=\sigma_3$. The doubled symmetry $\bar{Q}$ in Eq.~\ref{eq:doubsym} will then be diagonalized by the Hadamard-like matrix $U_Q=\frac{1}{\sqrt{2}}\begin{pmatrix}
\mathbb{I} & \mathbb{I}\\ \mathbb{I} & -\mathbb{I}
\end{pmatrix}$. Using the relation between type $Q$ symmetry sectors described in Sec.~\ref{sec:typeqsym}, we map
\begin{align}
\sigma_0\rightarrow -\sigma_{33},\quad \sigma_1\rightarrow\sigma_{22},\quad \sigma_2\rightarrow \sigma_{21},\quad \sigma_3\rightarrow\sigma_{30},
\end{align}
to obtain the complete Hamiltonian at the doubled level from the Hamiltonian in a single symmetry sector, in a basis where the type $Q$ symmetry is diagonalized. Conjugating by $U_Q$ to return to the usual basis, which maps the first element in the Pauli string $\sigma_1\rightarrow \sigma_3$, $\sigma_3\rightarrow \sigma_1$, $\sigma_2\rightarrow -\sigma_2$, and taking the upper right block of the doubled Hamiltonian, we find that the non-Hermitian Hamiltonian and symmetry implementations are
\begin{center}
\begin{tabular}{c|c|c|c}
Class & Hamiltonian & $C$ sym. & $Q$ sym.  \\
\hline
18 & $m_3\sigma_0+im_2\sigma_1-im_1\sigma_2$ & $\sigma_1$ & $\sigma_3$\\
\end{tabular}
\end{center}
The eigenvalue spectrum is $E=m_3\pm i\sqrt{m_1^2+m_2^2}$, where a positive value of $m_3$ can easily be deformed into a negative one simply by choosing a finite value for $m_1^2+m_2^2$ and varying $m_3$. Thus, all non-singular Hamiltonians belong to the same topological class in this example, consistent with our general results that gives a trivial classification in this case.

\bibliography{main}
\newpage

\begin{table*}\begin{center}\begin{tabular}{ | c | c | c | c | c | c | c | c | c |}\hline Sym. & Gen. Rel. & Clifford Generators & Cl. Sp. 0D & $d = 0$ & 1 & 2 & 3 & Known Class\\\hline1,  &  & $\{\gamma ,m,\Sigma \}$ &  $\mathcal{C}_1$ & $0$ & $\mathbb{Z}$ & $0$ & $\mathbb{Z}$ & Non-H A\\\hline2, P &  & $\{\gamma ,m,\Sigma \}_{\Sigma P}$ &  $\mathcal{C}_1^{\times 2}$ & $0$ & $\mathbb{Z}^{\times 2}$ & $0$ & $\mathbb{Z}^{\times 2}$ & Non-H AIII\\\hline3, Q &  & $\{\gamma ,m\}_{Q}$ &  $\mathcal{C}_0$ & $\mathbb{Z}$ & $0$ & $\mathbb{Z}$ & $0$ & A\\\hline4, PQ & $\epsilon_{pq}=1$ & $\{\gamma ,m,P\}_{Q}$ &  $\mathcal{C}_1$ & $0$ & $\mathbb{Z}$ & $0$ & $\mathbb{Z}$ & AIII\\\hline5, PQ & $\epsilon_{pq}=-1$ & $\{\gamma ,m\}_{Q,\Sigma P}$ &  $\mathcal{C}_0^{\times 2}$ & $\mathbb{Z}^{\times 2}$ & $0$ & $\mathbb{Z}^{\times 2}$ & $0$ & \\\hline6, C & $\epsilon_c=1$, $\eta_c=1$ & $\{\gamma ,Jm,\Sigma ,C,JC\}$ &  $\mathcal{R}_7$ & $0$ & $0$ & $0$ & $\mathbb{Z}$ & \\\hline7, C & $\epsilon_c=1$, $\eta_c=-1$ & $\{\gamma ,Jm,\Sigma ,C,JC\}$ &  $\mathcal{R}_3$ & $0$ & $\mathbb{Z}_2$ & $\mathbb{Z}_2$ & $\mathbb{Z}$ & \\\hline8, C & $\epsilon_c=-1$, $\eta_c=1$ & $\{J\gamma ,m,\Sigma ,C,JC\}$ &  $\mathcal{R}_3$ & $0$ & $\mathbb{Z}_2$ & $\mathbb{Z}_2$ & $\mathbb{Z}$ & \\\hline9, C & $\epsilon_c=-1$, $\eta_c=-1$ & $\{J\gamma ,m,\Sigma ,C,JC\}$ &  $\mathcal{R}_7$ & $0$ & $0$ & $0$ & $\mathbb{Z}$ & \\\hline10, PC & \begin{tabular}{c}$\epsilon_c=1$, $\eta_c=1$, $\epsilon_{pc}=1$\\\hline$\epsilon_c=-1$, $\eta_c=1$, $\epsilon_{pc}=1$\end{tabular} & \begin{tabular}{c}$\{\gamma ,Jm,\Sigma ,C,JC\}, [J\Sigma P]$\\\hline$\{J\gamma ,m,\Sigma ,C,JC\}, [J\Sigma P]$\end{tabular} &  $\mathcal{C}_1$ & $0$ & $\mathbb{Z}$ & $0$ & $\mathbb{Z}$ & \\\hline11, PC & \begin{tabular}{c}$\epsilon_c=1$, $\eta_c=-1$, $\epsilon_{pc}=1$\\\hline$\epsilon_c=-1$, $\eta_c=-1$, $\epsilon_{pc}=1$\end{tabular} & \begin{tabular}{c}$\{\gamma ,Jm,\Sigma ,C,JC\}, [J\Sigma P]$\\\hline$\{J\gamma ,m,\Sigma ,C,JC\}, [J\Sigma P]$\end{tabular} &  $\mathcal{C}_1$ & $0$ & $\mathbb{Z}$ & $0$ & $\mathbb{Z}$ & \\\hline12, PC & \begin{tabular}{c}$\epsilon_c=1$, $\eta_c=1$, $\epsilon_{pc}=-1$\\\hline$\epsilon_c=-1$, $\eta_c=-1$, $\epsilon_{pc}=-1$\end{tabular} & \begin{tabular}{c}$\{\gamma ,Jm,\Sigma ,C,JC\}_{\Sigma P}$\\\hline$\{J\gamma ,m,\Sigma ,C,JC\}_{\Sigma P}$\end{tabular} &  $\mathcal{R}_7^{\times 2}$ & $0$ & $0$ & $0$ & $\mathbb{Z}^{\times 2}$ & \\\hline13, PC & \begin{tabular}{c}$\epsilon_c=1$, $\eta_c=-1$, $\epsilon_{pc}=-1$\\\hline$\epsilon_c=-1$, $\eta_c=1$, $\epsilon_{pc}=-1$\end{tabular} & \begin{tabular}{c}$\{\gamma ,Jm,\Sigma ,C,JC\}_{\Sigma P}$\\\hline$\{J\gamma ,m,\Sigma ,C,JC\}_{\Sigma P}$\end{tabular} &  $\mathcal{R}_3^{\times 2}$ & $0$ & $\mathbb{Z}_2^{\times 2}$ & $\mathbb{Z}_2^{\times 2}$ & $\mathbb{Z}^{\times 2}$ & \\\hline14, QC & $\epsilon_c=1$, $\eta_c=1$, $\epsilon_{qc}=1$ & $\{\gamma ,Jm,C,JC\}_{Q}$ &  $\mathcal{R}_0$ & $\mathbb{Z}$ & $0$ & $0$ & $0$ & AI\\\hline15, QC & $\epsilon_c=1$, $\eta_c=-1$, $\epsilon_{qc}=1$ & $\{\gamma ,Jm,C,JC\}_{Q}$ &  $\mathcal{R}_4$ & $\mathbb{Z}$ & $0$ & $\mathbb{Z}_2$ & $\mathbb{Z}_2$ & AII\\\hline16, QC & $\epsilon_c=-1$, $\eta_c=1$, $\epsilon_{qc}=1$ & $\{J\gamma ,m,C,JC\}_{Q}$ &  $\mathcal{R}_2$ & $\mathbb{Z}_2$ & $\mathbb{Z}_2$ & $\mathbb{Z}$ & $0$ & D\\\hline17, QC & $\epsilon_c=-1$, $\eta_c=-1$, $\epsilon_{qc}=1$ & $\{J\gamma ,m,C,JC\}_{Q}$ &  $\mathcal{R}_6$ & $0$ & $0$ & $\mathbb{Z}$ & $0$ & C\\\hline18, QC & $\epsilon_c=1$, $\eta_c=1$, $\epsilon_{qc}=-1$ & $\{J\gamma ,m,\Sigma C,J\Sigma C\}_{Q}$ &  $\mathcal{R}_6$ & $0$ & $0$ & $\mathbb{Z}$ & $0$ & \\\hline19, QC & $\epsilon_c=1$, $\eta_c=-1$, $\epsilon_{qc}=-1$ & $\{J\gamma ,m,\Sigma C,J\Sigma C\}_{Q}$ &  $\mathcal{R}_2$ & $\mathbb{Z}_2$ & $\mathbb{Z}_2$ & $\mathbb{Z}$ & $0$ & \\\hline20, QC & $\epsilon_c=-1$, $\eta_c=1$, $\epsilon_{qc}=-1$ & $\{\gamma ,Jm,\Sigma C,J\Sigma C\}_{Q}$ &  $\mathcal{R}_4$ & $\mathbb{Z}$ & $0$ & $\mathbb{Z}_2$ & $\mathbb{Z}_2$ & \\\hline21, QC & $\epsilon_c=-1$, $\eta_c=-1$, $\epsilon_{qc}=-1$ & $\{\gamma ,Jm,\Sigma C,J\Sigma C\}_{Q}$ &  $\mathcal{R}_0$ & $\mathbb{Z}$ & $0$ & $0$ & $0$ & \\\hline\end{tabular}\end{center}\caption{\label{ref:tab1} First half of the periodic table with symmetries, commutation relations, Clifford algebra generators (a subscript indicates a commuting unitary symmetry, while a bracket indicates a unitary symmetry that squares to -1 and thus acts as an imaginary unit), classifying space, topological classification in low dimensions, as well as corresponding known classes.}\end{table*}

\begin{table*}\begin{center}\begin{tabular}{ | c | c | c | c | c | c | c | c | c |}\hline Sym. & Gen. Rel. & Clifford Generators & Cl. Sp. 0D & $d = 0$ & 1 & 2 & 3 & Known Class\\\hline22, PQC & \begin{tabular}{c}$\epsilon_c=1$, $\eta_c=1$, $\epsilon_{pq}=1$, $\epsilon_{pc}=1$, $\epsilon_{qc}=1$\\\hline$\epsilon_c=-1$, $\eta_c=1$, $\epsilon_{pq}=1$, $\epsilon_{pc}=1$, $\epsilon_{qc}=1$\end{tabular} & \begin{tabular}{c}$\{\gamma ,Jm,JP,C,JC\}_{Q}$\\\hline$\{J\gamma ,m,JP,C,JC\}_{Q}$\end{tabular} &  $\mathcal{R}_1$ & $\mathbb{Z}_2$ & $\mathbb{Z}$ & $0$ & $0$ & BDI\\\hline23, PQC & \begin{tabular}{c}$\epsilon_c=1$, $\eta_c=-1$, $\epsilon_{pq}=1$, $\epsilon_{pc}=1$, $\epsilon_{qc}=1$\\\hline$\epsilon_c=-1$, $\eta_c=-1$, $\epsilon_{pq}=1$, $\epsilon_{pc}=1$, $\epsilon_{qc}=1$\end{tabular} & \begin{tabular}{c}$\{\gamma ,Jm,JP,C,JC\}_{Q}$\\\hline$\{J\gamma ,m,JP,C,JC\}_{Q}$\end{tabular} &  $\mathcal{R}_5$ & $0$ & $\mathbb{Z}$ & $0$ & $\mathbb{Z}_2$ & CII\\\hline24, PQC & \begin{tabular}{c}$\epsilon_c=1$, $\eta_c=1$, $\epsilon_{pq}=-1$, $\epsilon_{pc}=1$, $\epsilon_{qc}=1$\\\hline$\epsilon_c=-1$, $\eta_c=1$, $\epsilon_{pq}=-1$, $\epsilon_{pc}=1$, $\epsilon_{qc}=-1$\\\hline$\epsilon_c=1$, $\eta_c=1$, $\epsilon_{pq}=-1$, $\epsilon_{pc}=1$, $\epsilon_{qc}=-1$\\\hline$\epsilon_c=-1$, $\eta_c=1$, $\epsilon_{pq}=-1$, $\epsilon_{pc}=1$, $\epsilon_{qc}=1$\end{tabular} & \begin{tabular}{c}$\{\gamma ,Jm,C,JC\}_{Q}, [J\Sigma P]$\\\hline$\{\gamma ,Jm,\Sigma C,J\Sigma C\}_{Q}, [J\Sigma P]$\\\hline$\{J\gamma ,m,\Sigma C,J\Sigma C\}_{Q}, [J\Sigma P]$\\\hline$\{J\gamma ,m,C,JC\}_{Q}, [J\Sigma P]$\end{tabular} &  $\mathcal{C}_0$ & $\mathbb{Z}$ & $0$ & $\mathbb{Z}$ & $0$ & \\\hline25, PQC & \begin{tabular}{c}$\epsilon_c=1$, $\eta_c=-1$, $\epsilon_{pq}=-1$, $\epsilon_{pc}=1$, $\epsilon_{qc}=1$\\\hline$\epsilon_c=-1$, $\eta_c=-1$, $\epsilon_{pq}=-1$, $\epsilon_{pc}=1$, $\epsilon_{qc}=-1$\\\hline$\epsilon_c=1$, $\eta_c=-1$, $\epsilon_{pq}=-1$, $\epsilon_{pc}=1$, $\epsilon_{qc}=-1$\\\hline$\epsilon_c=-1$, $\eta_c=-1$, $\epsilon_{pq}=-1$, $\epsilon_{pc}=1$, $\epsilon_{qc}=1$\end{tabular} & \begin{tabular}{c}$\{\gamma ,Jm,C,JC\}_{Q}, [J\Sigma P]$\\\hline$\{\gamma ,Jm,\Sigma C,J\Sigma C\}_{Q}, [J\Sigma P]$\\\hline$\{J\gamma ,m,\Sigma C,J\Sigma C\}_{Q}, [J\Sigma P]$\\\hline$\{J\gamma ,m,C,JC\}_{Q}, [J\Sigma P]$\end{tabular} &  $\mathcal{C}_0$ & $\mathbb{Z}$ & $0$ & $\mathbb{Z}$ & $0$ & \\\hline26, PQC & \begin{tabular}{c}$\epsilon_c=1$, $\eta_c=1$, $\epsilon_{pq}=1$, $\epsilon_{pc}=-1$, $\epsilon_{qc}=1$\\\hline$\epsilon_c=-1$, $\eta_c=-1$, $\epsilon_{pq}=1$, $\epsilon_{pc}=-1$, $\epsilon_{qc}=1$\\\hline$\epsilon_c=1$, $\eta_c=1$, $\epsilon_{pq}=1$, $\epsilon_{pc}=-1$, $\epsilon_{qc}=-1$\\\hline$\epsilon_c=-1$, $\eta_c=-1$, $\epsilon_{pq}=1$, $\epsilon_{pc}=-1$, $\epsilon_{qc}=-1$\end{tabular} & \begin{tabular}{c}$\{\gamma ,Jm,P,C,JC\}_{Q}$\\\hline$\{J\gamma ,m,P,C,JC\}_{Q}$\\\hline$\{J\gamma ,m,P,\Sigma C,J\Sigma C\}_{Q}$\\\hline$\{\gamma ,Jm,P,\Sigma C,J\Sigma C\}_{Q}$\end{tabular} &  $\mathcal{R}_7$ & $0$ & $0$ & $0$ & $\mathbb{Z}$ & CI\\\hline27, PQC & \begin{tabular}{c}$\epsilon_c=1$, $\eta_c=-1$, $\epsilon_{pq}=1$, $\epsilon_{pc}=-1$, $\epsilon_{qc}=1$\\\hline$\epsilon_c=-1$, $\eta_c=1$, $\epsilon_{pq}=1$, $\epsilon_{pc}=-1$, $\epsilon_{qc}=1$\\\hline$\epsilon_c=1$, $\eta_c=-1$, $\epsilon_{pq}=1$, $\epsilon_{pc}=-1$, $\epsilon_{qc}=-1$\\\hline$\epsilon_c=-1$, $\eta_c=1$, $\epsilon_{pq}=1$, $\epsilon_{pc}=-1$, $\epsilon_{qc}=-1$\end{tabular} & \begin{tabular}{c}$\{\gamma ,Jm,P,C,JC\}_{Q}$\\\hline$\{J\gamma ,m,P,C,JC\}_{Q}$\\\hline$\{J\gamma ,m,P,\Sigma C,J\Sigma C\}_{Q}$\\\hline$\{\gamma ,Jm,P,\Sigma C,J\Sigma C\}_{Q}$\end{tabular} &  $\mathcal{R}_3$ & $0$ & $\mathbb{Z}_2$ & $\mathbb{Z}_2$ & $\mathbb{Z}$ & DIII\\\hline28, PQC & \begin{tabular}{c}$\epsilon_c=1$, $\eta_c=1$, $\epsilon_{pq}=-1$, $\epsilon_{pc}=-1$, $\epsilon_{qc}=1$\\\hline$\epsilon_c=-1$, $\eta_c=-1$, $\epsilon_{pq}=-1$, $\epsilon_{pc}=-1$, $\epsilon_{qc}=-1$\end{tabular} & \begin{tabular}{c}$\{\gamma ,Jm,C,JC\}_{Q,\Sigma P}$\\\hline$\{\gamma ,Jm,\Sigma C,J\Sigma C\}_{Q,\Sigma P}$\end{tabular} &  $\mathcal{R}_0^{\times 2}$ & $\mathbb{Z}^{\times 2}$ & $0$ & $0$ & $0$ & \\\hline29, PQC & \begin{tabular}{c}$\epsilon_c=1$, $\eta_c=-1$, $\epsilon_{pq}=-1$, $\epsilon_{pc}=-1$, $\epsilon_{qc}=1$\\\hline$\epsilon_c=-1$, $\eta_c=1$, $\epsilon_{pq}=-1$, $\epsilon_{pc}=-1$, $\epsilon_{qc}=-1$\end{tabular} & \begin{tabular}{c}$\{\gamma ,Jm,C,JC\}_{Q,\Sigma P}$\\\hline$\{\gamma ,Jm,\Sigma C,J\Sigma C\}_{Q,\Sigma P}$\end{tabular} &  $\mathcal{R}_4^{\times 2}$ & $\mathbb{Z}^{\times 2}$ & $0$ & $\mathbb{Z}_2^{\times 2}$ & $\mathbb{Z}_2^{\times 2}$ & \\\hline30, PQC & \begin{tabular}{c}$\epsilon_c=-1$, $\eta_c=1$, $\epsilon_{pq}=-1$, $\epsilon_{pc}=-1$, $\epsilon_{qc}=1$\\\hline$\epsilon_c=1$, $\eta_c=-1$, $\epsilon_{pq}=-1$, $\epsilon_{pc}=-1$, $\epsilon_{qc}=-1$\end{tabular} & \begin{tabular}{c}$\{J\gamma ,m,C,JC\}_{Q,\Sigma P}$\\\hline$\{J\gamma ,m,\Sigma C,J\Sigma C\}_{Q,\Sigma P}$\end{tabular} &  $\mathcal{R}_2^{\times 2}$ & $\mathbb{Z}_2^{\times 2}$ & $\mathbb{Z}_2^{\times 2}$ & $\mathbb{Z}^{\times 2}$ & $0$ & \\\hline31, PQC & \begin{tabular}{c}$\epsilon_c=-1$, $\eta_c=-1$, $\epsilon_{pq}=-1$, $\epsilon_{pc}=-1$, $\epsilon_{qc}=1$\\\hline$\epsilon_c=1$, $\eta_c=1$, $\epsilon_{pq}=-1$, $\epsilon_{pc}=-1$, $\epsilon_{qc}=-1$\end{tabular} & \begin{tabular}{c}$\{J\gamma ,m,C,JC\}_{Q,\Sigma P}$\\\hline$\{J\gamma ,m,\Sigma C,J\Sigma C\}_{Q,\Sigma P}$\end{tabular} &  $\mathcal{R}_6^{\times 2}$ & $0$ & $0$ & $\mathbb{Z}^{\times 2}$ & $0$ & \\\hline32, PQC & \begin{tabular}{c}$\epsilon_c=1$, $\eta_c=1$, $\epsilon_{pq}=1$, $\epsilon_{pc}=1$, $\epsilon_{qc}=-1$\\\hline$\epsilon_c=-1$, $\eta_c=1$, $\epsilon_{pq}=1$, $\epsilon_{pc}=1$, $\epsilon_{qc}=-1$\end{tabular} & \begin{tabular}{c}$\{J\gamma ,m,JP,\Sigma C,J\Sigma C\}_{Q}$\\\hline$\{\gamma ,Jm,JP,\Sigma C,J\Sigma C\}_{Q}$\end{tabular} &  $\mathcal{R}_5$ & $0$ & $\mathbb{Z}$ & $0$ & $\mathbb{Z}_2$ & \\\hline33, PQC & \begin{tabular}{c}$\epsilon_c=1$, $\eta_c=-1$, $\epsilon_{pq}=1$, $\epsilon_{pc}=1$, $\epsilon_{qc}=-1$\\\hline$\epsilon_c=-1$, $\eta_c=-1$, $\epsilon_{pq}=1$, $\epsilon_{pc}=1$, $\epsilon_{qc}=-1$\end{tabular} & \begin{tabular}{c}$\{J\gamma ,m,JP,\Sigma C,J\Sigma C\}_{Q}$\\\hline$\{\gamma ,Jm,JP,\Sigma C,J\Sigma C\}_{Q}$\end{tabular} &  $\mathcal{R}_1$ & $\mathbb{Z}_2$ & $\mathbb{Z}$ & $0$ & $0$ & \\\hline34, K & $\eta_k=1$ & $\{\gamma ,Jm,J\Sigma ,K,JK\}$ &  $\mathcal{R}_1$ & $\mathbb{Z}_2$ & $\mathbb{Z}$ & $0$ & $0$ & Non-H AI/D\\\hline35, K & $\eta_k=-1$ & $\{\gamma ,Jm,J\Sigma ,K,JK\}$ &  $\mathcal{R}_5$ & $0$ & $\mathbb{Z}$ & $0$ & $\mathbb{Z}_2$ & Non-H AII/C\\\hline36, PK & $\eta_k=1$, $\epsilon_{pk}=1$ & $\{\gamma ,Jm,J\Sigma ,K,JK\}_{\Sigma P}$ &  $\mathcal{R}_1^{\times 2}$ & $\mathbb{Z}_2^{\times 2}$ & $\mathbb{Z}^{\times 2}$ & $0$ & $0$ & Non-H BDI\\\hline37, PK & $\eta_k=-1$, $\epsilon_{pk}=1$ & $\{\gamma ,Jm,J\Sigma ,K,JK\}_{\Sigma P}$ &  $\mathcal{R}_5^{\times 2}$ & $0$ & $\mathbb{Z}^{\times 2}$ & $0$ & $\mathbb{Z}_2^{\times 2}$ & Non-H CII\\\hline38, PK & \begin{tabular}{c}$\eta_k=1$, $\epsilon_{pk}=-1$\\\hline$\eta_k=-1$, $\epsilon_{pk}=-1$\end{tabular} & \begin{tabular}{c}$\{\gamma ,Jm,J\Sigma ,K,JK\}, [J\Sigma P]$\\\hline$\{\gamma ,Jm,J\Sigma ,K,JK\}, [J\Sigma P]$\end{tabular} &  $\mathcal{C}_1$ & $0$ & $\mathbb{Z}$ & $0$ & $\mathbb{Z}$ & Non-H CI/DIII\\\hline\end{tabular}\end{center}\caption{\label{ref:tab2} Second half of the periodic table with symmetries, commutation relations, Clifford algebra generators (a subscript indicates a commuting unitary symmetry, while a bracket indicates a unitary symmetry that squares to -1 and thus acts as an imaginary unit), classifying space, topological classification in low dimensions, as well as corresponding known classes.}\end{table*}

\end{document}